\documentclass[sigconf, nonacm]{acmart} 
\AtBeginDocument{%
  }

\usepackage{enumitem}
\usepackage{amsmath}
\usepackage{booktabs}
\usepackage{array}
\usepackage{caption}
\usepackage{subcaption}
\usepackage{siunitx}
\usepackage{multirow}
\usepackage{graphicx}
\usepackage{etoolbox}
\usepackage{longtable}
\usepackage{xcolor}
\usepackage{colortbl}
\usepackage{enumitem}
\usepackage{hyperref}
\usepackage{fontenc}
\usepackage{inputenc}
\usepackage{ragged2e}

\definecolor{headerblue}{RGB}{52, 90, 145}
\definecolor{rowgray}{RGB}{240, 244, 250}

\begin{document}

\title{OneSearch-V2: The Latent Reasoning Enhanced Self-distillation Generative Search Framework}

\author{Ben Chen$^{*\dagger}$, Siyuan Wang$^*$, Yufei Ma$^*$, Zihan Liang$^*$, Xuxin Zhang, Yue Lv, Ying Yang, Huangyu Dai, Lingtao Mao, Tong Zhao, Zhipeng Qian, Xinyu Sun, Zhixin Zhai, Yang Zhao, Bochao Liu, Jingshan Lv, Xiao Liang, Hui Kong, Jing Chen, Han Li, Chenyi Lei$^{\dagger}$, Wenwu Ou, Kun Gai}
\thanks{$\dagger$ Corresponding author. Homepage: \url{https://benchen4395.github.io/}}
\thanks{* Equal Contribution.}
\affiliation{
 \institution{Kuaishou Technology, Beijing, China}
 \country{Contact: \{benchen4395, leichenyi\}@gmail.com}
}

\renewcommand{\shortauthors}{Ben Chen et al.}

\begin{abstract}
Generative Retrieval (GR) has emerged as a promising paradigm for modern search systems. Compared to multi-stage cascaded architecture, it offers advantages such as end-to-end joint optimization and high computational efficiency. OneSearch, as a representative industrial-scale deployed generative search framework, has brought significant commercial and operational benefits. However, its inadequate understanding of complex queries, inefficient exploitation of latent user intents, and overfitting to narrow historical preferences have limited its further performance improvement.
To address these challenges, we propose \textbf{OneSearch-V2}, a latent reasoning enhanced self-distillation generative search framework. It contains three key innovations:
(1) a thought-augmented complex query understanding module, which enables deep query understanding and overcomes the shallow semantic matching limitations of direct inference; 
(2) a reasoning-internalized self-distillation training pipeline, which uncovers users' potential yet precise e-commerce intentions beyond log-fitting through implicit in-context learning; 
(3) a behavior preference alignment optimization system, which mitigates reward hacking arising from the single conversion metric, and addresses personal preference via direct user feedback.
Extensive offline evaluations demonstrate OneSearch-V2's strong query recognition and user profiling capabilities. Online A/B tests further validate its business effectiveness, yielding +3.98\% item CTR, +2.07\% buyer volume, and +2.11\% order volume. Manual evaluation  further confirms gains in search experience quality, with +1.37\% in page good rate and +1.65\% in query-item relevance. More importantly, OneSearch-V2 effectively mitigates common search system issues such as information bubbles and long-tail sparsity, without incurring additional inference costs or serving latency. Key codes are available at \url{https://github.com/benchen4395/onesearch-family}.
\end{abstract}



\maketitle

\begin{figure}[th]
  \centering
  \includegraphics[width=\linewidth]{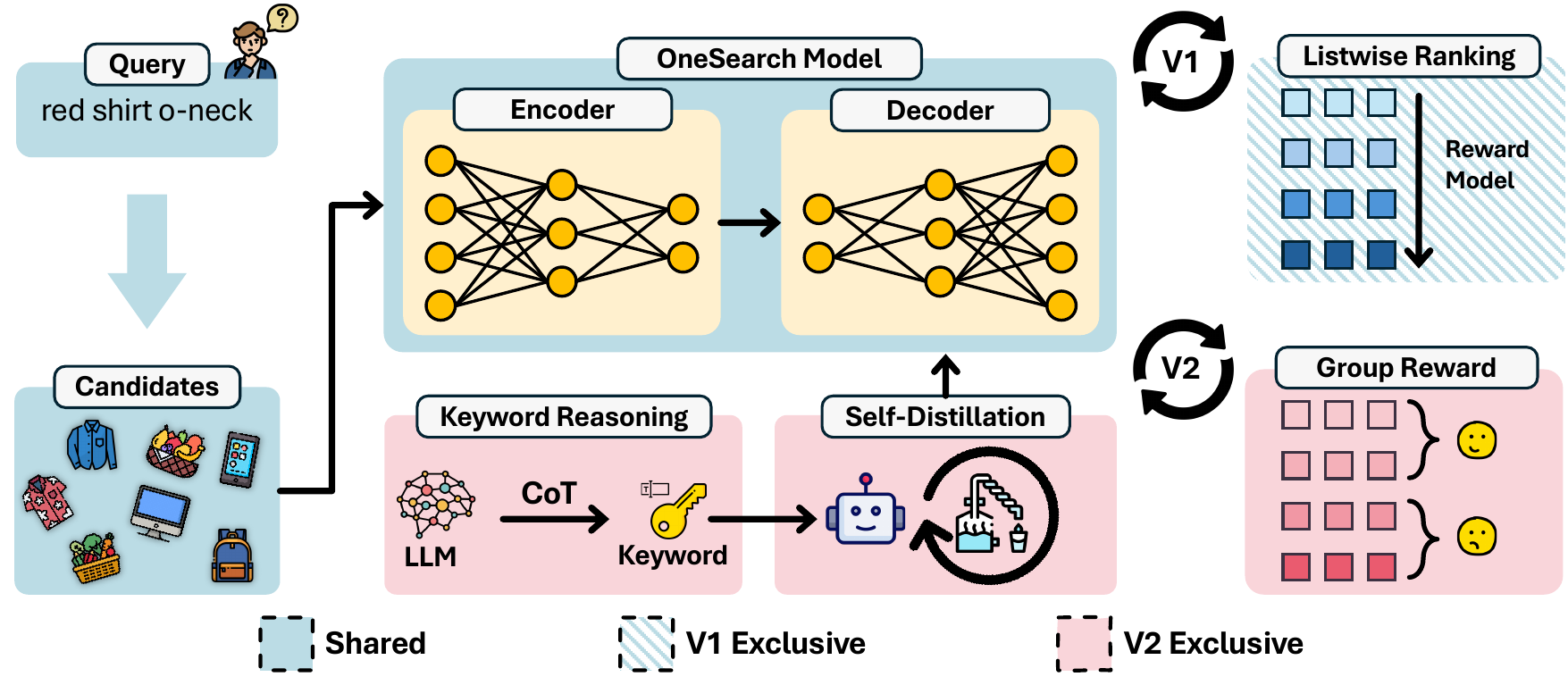}
  \caption{OneSearch-V2 vs.\ V1. OneSearch-V2 extends the generative search framework with thought-augmented query understanding, reasoning-internalized self-distillation, and behavior feedback preference alignment.}
  \label{Figure1}
\end{figure}

\section{Introduction}
Leveraging extensive world knowledge and powerful language modeling capabilities, large language models (LLMs) are profoundly reshaping search and recommendation systems. Compared to directly applying or distilling LLMs into smaller models for isolated modules such as recall, relevance, and ranking \cite{2024hstu,2025RankMixer,dong2026taosr1thinkingmodelecommerce,tang2025LREM}, a more promising research frontier lies in employing end-to-end generative retrieval to replace the traditional multi-stage cascading architecture, as exemplified by OneRec for video recommendation~\cite{2025onerec,zhou2025onerecv2technicalreport}, OneSug for query suggestion~\cite{2025onesug}, OneLoc for local life services~\cite{2025oneloc}, and MTGR for advertisement~\cite{Han_2025_MTGR}.
For e-commerce search, which requires jointly considering query-item relevance and user-item collaborative preference, the representative generative model is OneSearch~\cite{2025onesearch}. It enables direct optimization of the final objective, achieving superior business performance with substantially lower computational overhead.

\par
However, as user preferences grow increasingly diverse and search queries become more complex, we identify three key limitations that constrain the performance of OneSearch:

\par
1) \textit{Insufficient understanding of complex queries.} Typical search queries consist of 2--3 short keywords, yet many do not specify concrete item targets. For instance, ``indoor fitness equipment'' may reasonably correspond to treadmills or dumbbells, but not mountain bikes. Furthermore, long-tail queries frequently exhibit significant lexical disparity from target items, such as negation-type queries (e.g., ``relieve fatigue, no supplements'') and question-type queries (e.g., ``what swimming essentials?''). These complex queries demand stronger semantic understanding and reasoning capabilities. However, OneSearch takes the raw query as input and generates target items in a single forward pass under strict latency constraints. Although it incorporates category-level supervision during training (SFT Stage 1), the model still lacks the capacity for deep comprehension of these ambiguous queries.

\par
2) \textit{Insufficient personalized intent reasoning over user context.} Beyond query-level understanding, effective e-commerce search further requires reasoning over user-specific context---yet OneSearch's periodic updates rely heavily on historical co-occurrence patterns and log-fitting objectives, inevitably resulting in shallow matching that fails to uncover the true user intent. For example, given a user allergic to certain flowers searching for ``seasonal fresh flowers,'' the model should first reason about the current season, identify which varieties are in bloom, and proactively avoid allergenic species---even if such items historically exhibit strong conversion under the same query. While LLMs can excel at such personalized intent reasoning through explicit chain-of-thought (CoT), the substantially increased token generation renders test-time computation prohibitively expensive for online deployment.

\par
3) \textit{Fragile reward system with distributional bias.} The multi-stage, periodically updated preference reward system prevents OneSearch from adapting in a timely manner to newly emerging queries and user intents. Furthermore, the reward model, primarily trained on historical user behavior logs, is susceptible to inefficient sampling and potential reward hacking. These issues collectively cause OneSearch to overfit narrow historical preferences, reinforcing the long-tail distributional bias inherent in the search system.

\par
To bridge these gaps, we introduce \textbf{OneSearch-V2} a novel generative search framework enhanced by latent reasoning and self-distillation, shown in Fig.~\ref{Figure1}. It comprises three key contributions:
 
\par
1) \textbf{Thought-augmented query understanding module.} We employ LLMs to generate explicit CoT reasoning for each query-user pair, and construct compact keyword-based CoTs that maximize information density while emphasizing critical content (cf.~\cite{tang2025LREM}). 
These $\langle$query, user, CoTs$\rangle$ tuples serve as a semantic alignment corpus during training, enabling OneSearch to learn complex and personalized query interpretation. Moreover, the keyword-based CoTs can be directly injected into the model input as supplementary signals at inference time, yielding significant improvements for long-tail and ambiguous queries. They further serve as privileged teacher-side input for the latter self-distillation pipeline.

\par
2) \textbf{Reasoning-internalized self-distillation training pipeline.} We propose a self-distillation training mechanism to endow the generative model with latent reasoning capabilities, while avoiding the need for additional trainable parameters or special tokens as in existing latent reasoning methods~\cite{shen2025codi,hao2025coconut,zhang2026latentr3}. 
Instead, the reasoning ability is encoded into the model weights and internalized as intuition. To mitigate the representation instability caused by the information asymmetry between teacher and student in self-distillation, we jointly apply R-Drop for prediction consistency regularization and FGM for adversarial input robustness, with a unified forward pass design that reduces computational overhead.

\par
3) \textbf{Behavior feedback preference alignment optimization system.} We replace the previous hybrid ranking framework, which relies on a separately trained reward model, with a direct user interaction feedback optimization system. It leverages query-item relevance and user behavior signals as composite rewards, achieving an explicit trade-off between semantic matching and business conversion. We further introduce the SID overlap rate as an auxiliary reward for format validity and hierarchical content constraints. This design enables OneSearch-V2 to flexibly adjust reward composition according to various objectives, while also supporting streaming updates to handle newly emerging queries and intents in a timely manner.

\par
We conduct rigorous offline evaluations demonstrating that newer V2 achieves substantially higher recall and ranking performance than V1 across diverse complex e-commerce intents. Extensive online A/B tests on the Kuaishou mall search platform further confirm significant improvements: OneSearch-V2 achieves +3.98\% in item CTR, +1.17\% in PV CTR, +2.90\% in buyer volume, and +2.11\% in order volume, and a critical +3.45\% in GMV. Manual evaluations additionally reveal +1.37\% page good rate and +1.65\% query-item relevance, indicating meaningful gains in search experience quality. More importantly, OneSearch-V2 can be deployed without incurring additional inference cost or serving latency, while effectively mitigating information bubbles and long-tail sparsity without requiring a separate reward model. The codes and data case are publicly available at \url{https://github.com/benchen4395/onesearch-family}; we hope that open-sourcing this work will contribute to future advancements in generative retrieval.

\begin{figure*}[htbp]
    \centering
    \includegraphics[width=\textwidth]{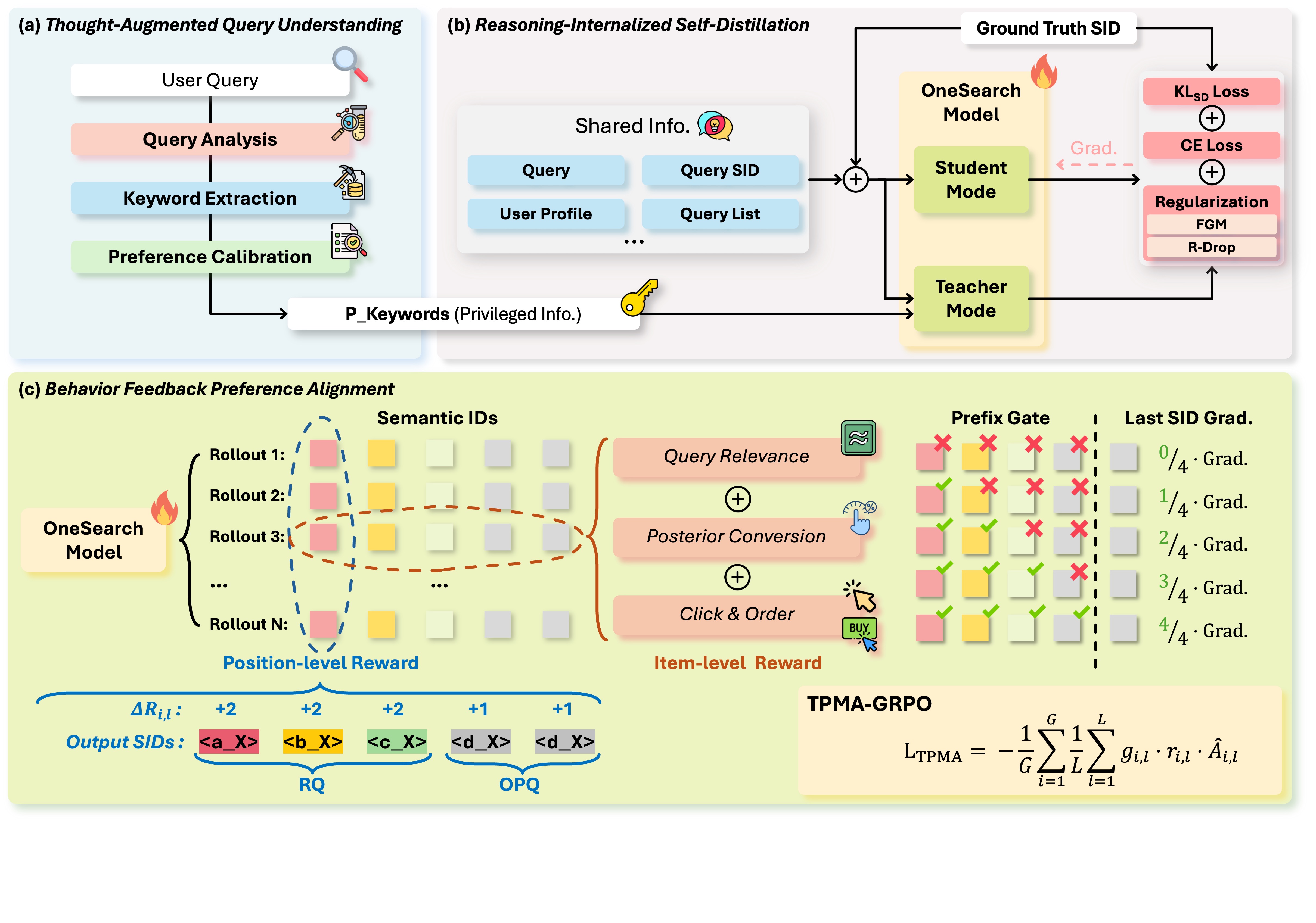}
    \vspace{-50pt}
    \caption{The Overall Framework of OneSearch V2. It contains (a) a thought-augmented complex query understanding module, (b) a reasoning-internalized self-distillation training pipeline, and (c) a behavior preference alignment optimization system. OneSearch-V2 effectively mitigates common search system issues such as information bubbles and long-tail sparsity, without incurring additional inference costs or serving latency.}
    \label{fig:method}
\end{figure*}

\section{Related Works}
\subsection{Generative Retrieval and Recommendation}
Generative retrieval and recommendation reframe item retrieval as a sequence generation problem, where a model directly produces discrete Semantic IDs (SIDs) of items in an autoregressive manner.
TIGER~\cite{2023tiger} is the seminal work in this line: it employs a Residual Quantized Variational Autoencoder (RQ-VAE) to compress item content embeddings into hierarchical discrete token sequences, and trains a Transformer-based encoder-decoder to autoregressively predict the next item's SID given a user's interaction history, elegantly unifying semantic content knowledge with collaborative signals through a shared quantized vocabulary.
By contrast, IDGenRec~\cite{tan2024idgenrec} represents items by a \emph{textual} ID generator natively in the LLM vocabulary, bridging the semantic gap between generative models and item ID spaces, and further enabling stronger cross-domain transfer.

\par
As these generative models scale to billion-parameter LLMs, their inference capabilities are greatly enhanced, but inference latency also becomes a critical bottleneck in deployment.
Lin et al.~\cite{lin2025efficientinfer} addressed this by adapting speculative decoding to LLM-based generative recommenders, where a lightweight draft model proposes candidate SID sequences that the large target model verifies in parallel, yielding significant speedup with negligible quality loss.
EARN~\cite{yang2025earn} further proposed inserting compact register tokens into the input sequence to cache repetitive intermediate computations across decoding steps, reducing inference latency for deployment-scale systems.
Most recently, R\textsuperscript{2}ec~\cite{you2025r2ec} introduced the first unified architecture that intrinsically integrates a reasoning chain into the generative recommendation loop: by optimizing both reasoning and recommend head, it achieves substantial gains on diverse scenarios with competitive inference efficiency. These advances collectively motivate the core question of our work: how to equip generative search models with reasoning capabilities while keeping inference cost practical for online deployment.
 
\subsection{Latent Reasoning and Self-Distillation}
Explicit chain-of-thought (CoT) reasoning has proven effective in enhancing LLM performance on complex tasks~\cite{2022cot,2023tot,2024deepseekmath}, yet the increased token generation incurs prohibitive latency for online deployment. To circumvent this cost, latent reasoning methods internalize reasoning into continuous hidden representations without explicit verbalization. Coconut~\cite{hao2025coconut} replaces textual reasoning steps with continuous thought vectors in the latent space. CODI~\cite{shen2025codi} compresses explicit CoT into continuous thought vectors through single-stage self-distillation with L1 hidden-state alignment. Latent-R3~\cite{zhang2026latentr3} further applies reinforcement learning over continuous latent representations. While these methods avoid explicit CoT at inference time, they typically require architectural modifications such as additional token embeddings, projection layers, or specialized decoding, which complicate deployment.

\par
A closely related line of work achieves reasoning internalization through \textit{information-asymmetric self-distillation}, where the same model serves as both teacher and student, with the asymmetry arising from different input contexts rather than distinct parameter sets. SDFT~\cite{shenfeld2026selfdistillationenablescontinuallearningSDFT} constructs this asymmetry through in-context learning: the teacher observes few-shot demonstrations while the student sees only the raw query, and alignment is performed via reverse KL divergence. 
OPSD~\cite{zhao2026selfdistilledreasoneronpolicyselfdistillationOPSD} applies a similar paradigm to mathematical reasoning with reference-solution-augmented teachers and token-level JSD alignment. 
SDPO~\cite{SDPO} further extends this to RL settings, using environment feedback as privileged teacher information to construct dense per-token advantages from the teacher-student logit gap. These methods share a common insight: rich supervision can be extracted from the logit discrepancy between information-advantaged and information-deprived views of the same model, without any external teacher.
Our work extends this paradigm to generative retrieval for e-commerce search, where the output space shifts from natural language to discrete SID sequences and the information asymmetry is constructed through keyword-based CoTs derived from query understanding.

\subsection{Preference Alignment for GRs}
Reinforcement learning (RL) has been extensively explored to align generative retrieval and recommendation models with complex user preferences. 
OneRec~\cite{2025onerec} introduces Early Clipped GRPO (ECPO) to optimize a personalized Preference Score derived from a separately trained multi-objective reward model. To mitigate the instability and reward hacking issues associated with reward models, OneRec-V2~\cite{zhou2025onerecv2technicalreport} leverages Gradient-Bounded Policy Optimization (GBPO) directly on real-world user feedback signals, such as duration-aware watch time. 
Similarly, OneSug~\cite{2025onesug} adopts a reward-weighted ranking strategy based on fine-grained behavior-level weights. 

\par
Despite these advancements, existing RL methods share two critical limitations. 
First, methods relying on separately trained reward models are susceptible to sampling bias and reward hacking, as these models tend to overfit to a narrow subset of historical logs that only approximate global behavior distributions. 
Second, GRPO and its variants (e.g., ECPO, GBPO) assign a uniform, sequence-level advantage to every token within a generated SID sequence. 
However, SID generation follows a strict hierarchical causal structure, progressing from coarse-grained categories to fine-grained item attributes. 
Under this structure, a correct prefix followed by error suffixes has fundamentally different implications from entirely incorrect tokens.
Uniform advantage assignment conflates these distinct positional contributions, weakening the learning signal for fine-grained token generation.
Our work addresses both limitations: we replace the separate reward model with direct behavior feedback and introduce a token-position marginal advantage mechanism that respects the hierarchical nature of SID sequences.

\section{Methodology}
In this section, we detail OneSearch-V2, the latent reasoning enhanced self-distillation generative search framework. First, we explore whether multimodal or unimodal SID tokenization is more suitable for e-commerce generative retrieval in \S~\ref{Unimodal-SID}. We then introduce the thought-augmented query understanding module in \S~\ref{TAQU}, and elaborate on the reasoning-internalized self-distillation training pipeline in \S~\ref{self-distill}. Finally, in \S~\ref{GRPO_opt}, we propose the behavior feedback preference alignment optimization system, which directly adopts user interaction feedback to replace multiple reward models for personalized ranking. The overall framework is shown in Fig.~\ref{fig:method}.

\subsection{Multimodal or Unimodal SID Tokenization?}
\label{Unimodal-SID}
Semantic IDs (SIDs) have emerged as a cornerstone for GR systems due to their efficient and hierarchical semantic representation. Extensive research has investigated efficient SID tokenization~\cite{lixiaopeng2025genrecsurvey, jia2025principlesapplicationscomprehensivesurvey}, which can be broadly categorized into two types: unimodal and multimodal. Here we explore which encoding paradigm is more suitable for e-commerce generative search.

\par
Unlike recommendation systems, search engines must address the critical challenge of aligning queries and items within a unified tokenization to ensure robust semantic constraints. 
This necessitates careful handling of the representational disparity between unimodal queries and multimodal item contents, as items are characterized by extensive textual descriptions, multiple images showing different perspectives, and even explanatory videos. 
OneSearch-V1 addresses this by transforming multimodal information into a unimodal representation. 
Specifically, it employs Qwen-VL~\cite{Qwen-VL} to extract core keywords from diverse sources, thereby constructing a unified textual representation. 
Alternative approaches adopt direct multimodal mapping, either by feeding all sources simultaneously into the model or by encoding individual modalities separately before concatenation. 
However, these methods face inherent limitations: multiple images may display mutually exclusive attributes (e.g., a dress available in different colors), and the abundance of redundant attributes may introduce extra bias (e.g., number and position of T-shirt buttons). Consequently, core attributes risk being obscured in the multimodal encoding process.

\par
To comprehensively compare the effectiveness of multimodal versus unimodal tokenization, we conducted experiments across multiple model configurations, including: a) Unimodal encoding utilizing text descriptions only, b) Multimodal encoding, containing unified encoding (joint processing) and separate encoding with subsequent concatenation, as well as c) Keyword hierarchical quantization in OneSearch. For experimental simplicity, we collected about 5M online clicked <query, item> pairs, and restricted the item input to only the title and two primary pictures. All embeddings were subsequently tokenized using the unified RQ-OPQ framework. The results are depicted in Table~\ref{tab:encoding_comparison}.

\begin{table}[!t]
\centering
\small
\caption{Comparison of unimodal, multimodal, and KHQE tokenization approaches. Recall@10 and MRR@10 are evaluated on click data.}
\label{tab:encoding_comparison}
\begin{tabular}{llccccc}
\toprule
\textbf{Type} & \textbf{Model}$^*$ & \textbf{Size} & \textbf{CUR} & \textbf{ICR} & \textbf{Recall} & \textbf{MRR} \\
\midrule
\multirow{2}{*}{uni-} & bge-base & 109M & 4.54\% & 96.88\% & 0.2445 & 0.1013 \\
 & qwen3 & 0.6B & \textbf{5.11\%} & \underline{97.56\%} & \underline{0.2468} & \underline{0.1025} \\
\midrule
\multirow{4}{*}{multi-} & uniecs & 200M & 4.54\% & 94.62\% & 0.2368 & 0.1007 \\
 & bge-vl & 149M & 4.23\% & 94.46\% & 0.2364 & 0.1009 \\
 & qwen3-vl & 2B & 4.86\% & 95.27\% & 0.2389 & 0.1012 \\
 & CLIP & 188M & 4.03\% & 94.16\% & 0.2358 & 0.1003 \\
\midrule
KHQE & bge+kw. & 109M & \textbf{5.11\%} & \textbf{99.50\%} & \textbf{0.2542} & \textbf{0.1085} \\
\bottomrule
\end{tabular}
\begin{flushleft}
\small
$^*$Bge-base and bge-vl are from~\cite{xiao2024BGE}, qwen3 and qwen3-vl from~\cite{yang2025qwen3,bai2025qwen3vltechnicalreport}, uniecs is the cross-modal retrieval model~\cite{liang2025uniecs}, and CLIP is from~\cite{ilharco_gabriel_openclip}.
\end{flushleft}
\end{table}

\par
Unimodal approaches significantly outperform multimodal counterparts at comparable scales—even the smaller bge-base surpasses the larger Qwen3-VL. This gap stems from cross-modal representational discrepancies and redundant attributes that constrain multimodal encoding effectiveness. The separate-then-concatenate strategy performs worst, further confirming these challenges. KHQE achieves optimal results, demonstrating superior core attribute extraction and hierarchical representation. More importantly, its smaller size allows real-time processing of input queries, striking a favorable balance between performance and efficiency. Meanwhile, these findings underscore that developing discriminative encodings for e-commerce search should highlight two critical factors: mitigating cross-modal disparities and enhancing salient information.

\subsection{Thought-Augmented Query Understanding}
\label{TAQU}
\begin{figure}[!t]
  \centering
  \includegraphics[width=\linewidth]{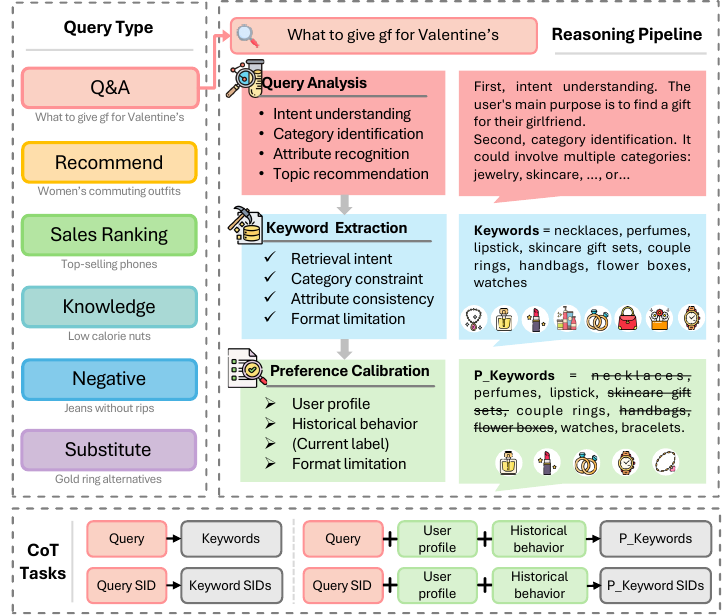}
  \caption{  Three-step keyword-based CoT extraction pipeline for diverse complex query types, along with the corresponding CoT tasks.}
  \label{fig:cot}
\end{figure}

E-commerce search engines handle massive volumes of queries exhibiting complex and heterogeneous user intents on a daily basis, including: (1) \textit{head queries} characterized by highly-divergent and underspecified intent (e.g., ``indoor fitness equipment''); and (2) \textit{tail queries} that encompass diverse types (see Fig.~\ref{fig:cot}), imposing intricate semantic constraints. On the Kuaishou Mall platform, these complex queries constitute about one-third of total page views (PV) yet account for merely 8\% of conversions, indicating a disproportionately low conversion rate. While OneSearch-V1 \cite{2025onesearch} partially alleviates the semantic discrepancy between complex queries and candidate items through aligned and enhanced representations, it remains fundamentally constrained by the inherent difficulties at both ends of the query frequency distribution, as evidenced by an inverted-U pattern of CTR gains: lower for head and tail queries, while higher for torso. The bottlenecks are fundamentally different: for head queries, the abundant co-occurrence patterns in interaction logs induce a broad candidate space, confronting the model with a \textit{``which to retrieve''} dilemma; for tail queries, the diverse and heterogeneous query formulations make intent understanding and item matching substantially harder, and the scarcity of behavioral signals further compounds this challenge, leaving the model in a \textit{``what can be retrieved''} dilemma.

\par
The emergence of explicit chain-of-thought (CoT) reasoning has enabled LLMs to achieve transparent and verifiable reasoning pathways for a wide range of complex tasks \cite{2024got,2024deepseekmath,2022cot,2023tot}. This advancement has inspired us to leverage CoT to address the query semantic dilemma. However, full and unconstrained CoT reasoning, which prioritizes mimicking human expression patterns, typically produces excessively lengthy outputs that small-scale models cannot efficiently generate. The heterogeneous nature between item SIDs and textual CoTs further obstructs straightforward inference. Moreover, e-commerce systems often require focusing solely on key terms aligned with query intent rather than comprehensive reasoning chains. These limitations motivate us to consider how to implement semantically-enhanced reasoning more efficiently.

\par
Here we propose a thought-augmented query understanding schema. 
We first employ LLMs to generate precise CoTs governed by four progressive constraints, and then extract keyword sets of elevated information density, subject to intent, category, and attribute consistency. 
Unlike a recent work on reasoning-then-embedding dense retrieval \cite{tang2025LREM}, our method demonstrates superior capability in extracting high e-commerce intent queries, excluding queries with non-intent and provides more explicit predictions regarding latent attribute preferences. These extracted keywords serve as supplementary semantic signals during training to enhance query intent recognition and user preference calibration.

\subsubsection{Keyword-based CoT}
\label{cot}
This paradigm circumvents the excessive computational overhead incurred by lengthy CoTs during inference, which confer merely marginal information utility. 
As shown in Fig.~\ref{fig:cot}, we formulate a three-step reasoning pipeline, with detailed prompt templates provided in Appendix~\ref{prompt}:
\begin{enumerate}[leftmargin=*, label=\textbf{\arabic*.}]
\item \textbf{Query Analysis.} We formulate an analysis scheme comprising four components.
(i) \textit{Intent understanding}, which identifies the primary retrieval target (i.e., merchandise, shop, or live-stream anchor); 
(ii) \textit{category identification}, which performs hierarchical category matching from coarse to fine granularity; 
(iii) \textit{attribute recognition}, which extracts the attribute type and its corresponding value from the query; 
and (iv) \textit{topic recommendation}, which speculates potential candidate topics satisfying the user's need.

\item \textbf{Keyword Extraction.} For queries with merchandise retrieval intent, we extract keywords from the full analysis, subject to constraints on intent, category, and attribute consistency. 
The extracted keywords are subsequently refined through synonym merging and redundant word removal, and finally ranked in descending order of item popularity. 
For queries with other intent types, which are handled by dedicated matching engines, the pipeline terminates directly.

\item \textbf{Preference Calibration.}
Leveraging the user profile and historical behavioral signals, such as previously entered queries and interacted item sequences, the LLM perceives user preferences and filters or augments the extracted keyword set to better align with individual interests. 
During training, the items interacted within the current session are further injected as signals, thereby ensuring that keywords associated with ground-truth items are either preserved or explicitly introduced into the set.

\end{enumerate}

\subsubsection{Training Paradigm Refinement}
\label{training paradigm}
The resulting $\langle$\textit{query}, \textit{keywords}$\rangle$ tuples from Step~2 and $\langle$\textit{query}, \textit{user}, \textit{keywords}$\rangle$ tuples from Step~3 collectively constitute the training corpus. We then incorporate four CoT tasks (shown in Fig.~\ref{fig:cot}) into the SFT Stage~1 (Semantic Alignment Procedure) of OneSearch-V1. As shown in Table~\ref{tab:train},\ref{tab:COT_result},\ref{tab:ablation_head_tail_query}, these tasks ($\backslash$+\ CoT tasks) guide the model to acquire richer query knowledge beyond historical logs and explore preference-aware item topics, thereby instructing the model to engage in more complex and personalized reasoning.

\par
During online deployment, the entire keyword-based CoT generation process for each distinct query is performed asynchronously and then used for streaming training and near-line inference. For the same query or $\langle$query, user$\rangle$ pair, cached, already computed content can be reused directly. This minimizes computational overhead and does not impact online inference latency.

\begin{table*}[t]
\centering
\caption{The overall training procedure of OneSearch-V2. It contains a three-stage supervised fine-tuning schema for semantic alignment, co-occurrence synchronization, and user personalization modeling, followed by a direct behavior feedback preference alignment for personalized preference learning.}
\begin{tabular}{lcccc}
\toprule
\textbf{Procedure} & \textbf{SFT Stage 1} & \textbf{SFT Stage 2} & \textbf{SFT Stage 3} & \textbf{RL Stage} \\
\midrule
\textbf{Objective} 
& Semantic alignment  
& $\langle\text{\(q\), \(i\)}\rangle \ co\text{-}occurrence$
& User personalization
& Preference Alignment \\
\hline
\textbf{Component} 
&$\begin{array}{c}
\text{query/item} \leftrightarrow \text{SID} \\
\text{query/item} \mapsto \text{category} \\ 
\text{SID} \mapsto \text{category} \\
\text{\textbf{CoT}}\ \text{\textbf{tasks}}
\end{array}$
&$\begin{array}{c}
\text{query} \leftrightarrow \text{item} \\
\text{$SID_q$} \leftrightarrow \text{$SID_i$} \\
\end{array}$
& $\begin{bmatrix} uid\ \&\ q\\ {SID}_q\ \& \ Seq_q \\ \text{\(Seq_{short}\)}\ \& \ \text{\(Seq_{long}^{emb}\) } \\ \text{\(keywords\)} \ \text{(\textbf{RAG})} \end{bmatrix} \mapsto {SID}_q$
& $\begin{bmatrix} \text{user \& query} \\ \text{seq. feat.} \\ \text{$item_{clk/order}$} \\ \text{$item_{rollout}$} \end{bmatrix} \mapsto \text{Rank Score}$ \\
\bottomrule
\label{tab:train}
\end{tabular}
\end{table*}

\begin{table}[htbp]
\centering
\caption{Results of CoT task augmentation and keyword injection as information gains, where $n = 10$.}
\begin{tabular}{lcccc}
\toprule
\multirow{2}{*}{\textbf{Model}} 
  & \multicolumn{2}{c}{\textbf{Order (7229)}}
  & \multicolumn{2}{c}{\textbf{Click (30k)}} \\
\cmidrule(lr){2-3} \cmidrule(lr){4-5}
  & \textbf{HR@n} & \textbf{MRR@n} 
  & \textbf{HR@n} & \textbf{MRR@n} \\
\midrule
baseline
  & 0.2046 & 0.0985 
  & 0.2231 & 0.0728  \\
$\backslash$+\ CoT tasks
  & 0.2094 & 0.1008 
  & 0.2266 & 0.0731 \\
\midrule
+ direct CoT
  & 0.0898 & 0.0189 
  & 0.1013 & 0.0146  \\
+ RAG
  & \textbf{0.2139} & \textbf{0.1011} 
  & \textbf{0.2327} & \textbf{0.0743}  \\
\bottomrule
\end{tabular}
\label{tab:COT_result}
\end{table}

\begin{table}[htbp]
\centering
\caption{Ablation study of CoT task augmentation on head and tail query types, where $n = 10$.}
\label{tab:ablation_head_tail_query}
\begin{tabular}{lcccc}
\toprule
\multirow{2}{*}{\textbf{Model}} 
  & \multicolumn{2}{c}{\textbf{Head}} 
  & \multicolumn{2}{c}{\textbf{Tail}} \\
\cmidrule(lr){2-3} \cmidrule(lr){4-5}
  & \textbf{HR@n} & \textbf{MRR@n} 
  & \textbf{HR@n} & \textbf{MRR@n} \\
\midrule
baseline
  & 0.2362 & 0.0817 
  & 0.1952 & 0.0733  \\
$\backslash$+\ CoT tasks
  & 0.2419 & 0.0829 
  & 0.1963 & 0.0734 \\
\midrule
+ direct CoT
  & 0.1116 & 0.0180 
  & 0.0809 & 0.0120 \\
+ RAG
  & \textbf{0.2438} & \textbf{0.0845} 
  & \textbf{0.1973} & \textbf{0.0779}  \\
\bottomrule
\end{tabular}
\end{table}

\subsection{Reasoning-Internalized Self-Distillation}
\label{self-distill}
An intuitive approach would be to train OneSearch to first generate reasoning keywords, followed by candidate SIDs. However, the representational heterogeneity between discrete SIDs and textual keywords poses a severe challenge for small-scale generative models. As demonstrated in Table~\ref{tab:COT_result}-\ref{tab:ablation_head_tail_query}, explicit CoT reasoning (+ direct CoT) substantially degrades OneSearch's performance, yielding results even significantly inferior to even the baseline. Instead, we leverage these keywords as supplementary information for queries at the input layer, as shown in Fig.~\ref{fig:method}. Notably, this input-augmented method (+ RAG) further enhances the model's retrieval and ranking effectiveness.

\par
However, obtaining these CoTs at inference time incurs non-trivial latency overhead, as it requires an additional call to the thought-augmented query understanding module per request, which is prohibitive under the strict latency constraints of online e-commerce search. Moreover, the limited coverage of keyword-based CoTs may also restrict the model to inferring only items explicitly covered by the keyword set. For example, when a user searches for "hotel essentials," if the keywords are limited to towels, toothbrushes, and razors, OneSearch might fail to recommend disposable slippers. These challenges raise a fundamental question: \textit{can we retain or even further enhance the performance gains of reasoning without bearing its inference cost?}

\par
We address this by proposing a \textbf{reasoning-internalized self-distillation} mechanism that transfers the explicit reasoning capability into the model's parameters, effectively converting deliberate, keyword-guided CoTs into fast, intuition-like inference. 
Unlike prior latent reasoning approaches that introduce additional trainable tokens~\cite{hao2025coconut} or continuous thought vectors~\cite{shen2025codi,zhang2026latentr3} into the decoding process, our method requires \textit{no architectural modification, no extra parameters, and no additional inference tokens}. The reasoning ability is encoded entirely into the existing model weights through a carefully designed distillation pipeline.

\subsubsection{Self-Distillation Formulation}
\par
Our self-distillation operates on the principle of \textit{information asymmetry}: the teacher observes strictly richer input than the student, while the student is trained to match the teacher's output distribution despite this informational disadvantage. Crucially, the teacher and student share the same model weights, eliminating the need for a separate teacher network and halving the memory footprint compared to conventional knowledge distillation.

\par
Concretely, let $\mathcal{M}_\theta$ denote the generative model parameterized by $\theta$. For a given training sample, the teacher receives the full input prompt augmented with keyword-based CoTs from \S\ref{cot}:
\begin{equation}
    x^{(T)} = (\text{uid},\, q,\, \text{SID}_q,\, \text{Seq}_q,\, \text{Seq}_{\text{short}},\, \text{Seq}_{\text{long}}^{\text{emb}},\, \textbf{kw}),
\end{equation}
where $\textbf{kw}$ denotes the personalized keyword-based CoTs. The student receives the same prompt \textit{without} the keyword augmentation:
\begin{equation}
    x^{(S)} = (\text{uid},\, q,\, \text{SID}_q,\, \text{Seq}_q,\, \text{Seq}_{\text{short}},\, \text{Seq}_{\text{long}}^{\text{emb}}).
\end{equation}
Both the teacher and the student produce output logits over the target label sequence $y = (y_1, \dots, y_L)$:
\begin{equation}
    z^{(T)} = \mathcal{M}_\theta(y \mid x^{(T)}), \quad z^{(S)} = \mathcal{M}_\theta(y \mid x^{(S)}).
\end{equation}
Since $\theta$ is shared, the difference between $z^{(T)}$ and $z^{(S)}$ arises solely from the presence or absence of keyword information in the input. The distillation objective encourages the student to close this gap:
\begin{equation}
    \mathcal{L}_{\text{KL}} = \frac{1}{|\mathcal{V}|} \sum_{t \in \mathcal{V}} \text{KL}\!\Big(\text{softmax}\!\big(z_t^{(T)} / \tau\big) \;\Big\|\; \text{softmax}\!\big(z_t^{(S)} / \tau\big)\Big) \cdot \tau^2,
    \label{eq:kl_distill}
\end{equation}
where $\mathcal{V} = \{t : y_t \neq -100\}$ is the set of valid (non-padding) token positions, and $\tau$ is the distillation temperature. The teacher's logits are detached from the computational graph so that the KL gradient updates only the student's forward path. During training, the teacher forward pass is executed under \texttt{torch.no\_grad()}, and only the student path accumulates gradients.

\par
The base training objective for the student combines the standard cross-entropy loss with the distillation signal:
\begin{equation}
    \mathcal{L}_{\text{base}} = \mathcal{L}_{\text{CE}}(z^{(S)}, y) + \alpha_{\text{KL}} \cdot \mathcal{L}_{\text{KL}},
    \label{eq:base_sdft}
\end{equation}
where $\alpha_{\text{KL}}$ controls the relative strength of the distillation signal.

\subsubsection{Mitigating Representation Instability}

\par
The information asymmetry between teacher and student introduces a fundamental challenge: the student must produce equally confident predictions from strictly less informative inputs. This forces the model's loss surface to become sharper in the neighborhood of keyword-absent inputs, as small perturbations in the embedding space can cause disproportionately large changes in the output distribution. We identify two complementary failure modes and address each with a targeted regularization technique.

\paragraph{Prediction Consistency via R-Drop.}
When the student lacks keyword guidance, its internal representations for semantically ambiguous queries become sensitive to stochastic perturbations from dropout. Two forward passes of the same input through the student may yield inconsistent output distributions, indicating that the model has not robustly internalized the query semantics. To enforce prediction stability, we apply R-Drop regularization~\cite{NEURIPS2021Rdrop}, which performs two forward passes $z_1^{(S)}$, $z_2^{(S)}$ with independent dropout masks and minimizes their divergence:
\begin{equation}
    \mathcal{L}_{\text{R-Drop}} = \frac{1}{2}\Big[\text{KL}(P_1 \| P_2) + \text{KL}(P_2 \| P_1)\Big],
    \label{eq:rdrop}
\end{equation}
where $P_k = \text{softmax}(z_k^{(S)})$ for $k \in \{1, 2\}$, and the KL terms are masked to valid token positions. This symmetric penalty discourages the model from relying on fragile internal pathways that are sensitive to dropout noise.

\paragraph{Input Robustness via Adversarial Perturbation.}
Complementary to R-Drop's output-space regularization, we apply Fast Gradient Method (FGM)~\cite{miyato2021fgm} to regularize the input embedding space. After the first backward pass, FGM perturbs the shared embedding layer along its gradient direction:
\begin{equation}
    r_{\text{adv}} = \epsilon \cdot \frac{\nabla_{e} \mathcal{L}_{\text{base}}}{\|\nabla_{e} \mathcal{L}_{\text{base}}\|_2},
    \label{eq:fgm}
\end{equation}
where $e$ denotes the embedding parameters, $\epsilon$ controls the perturbation magnitude, and $\mathcal{L}_{\text{base}} = \mathcal{L}_{\text{CE}} + \alpha_{\text{KL}} \cdot \mathcal{L}_{\text{KL}} + \alpha_{\text{R}} \cdot \mathcal{L}_{\text{R-Drop}}$. A second forward--backward pass on the perturbed embeddings $e + r_{\text{adv}}$ yields $\mathcal{L}_{\text{adv}}$, whose gradients are accumulated before restoring $e$. This smooths the loss landscape around each input, preventing sharp decision boundaries in regions where neighboring embeddings may correspond to semantically distinct queries.
\subsubsection{Total Optimization Objective}
Combining all components, the student objective is:
\begin{equation}
    \mathcal{L}_{\text{SDFT}} = \mathcal{L}_{\text{CE}} + \alpha_{\text{KL}} \cdot \mathcal{L}_{\text{KL}} + \alpha_{\text{R}} \cdot \mathcal{L}_{\text{R-Drop}} + \mathcal{L}_{\text{adv}},
    \label{eq:total_sdft}
\end{equation}
where $\mathcal{L}_{\text{adv}}$ denotes the cross-entropy and weighted distillation losses on the perturbed input (reusing $\alpha_{\text{KL}}$). We further replace standard cross-entropy with focal loss~\cite{lin2018focallossdenseobject} to mitigate the long-tail class imbalance in the SID vocabulary.

\subsection{Behavior Feedback Preference Alignment}
\label{GRPO_opt}
OneSearch-V1 adopts a hybrid ranking framework in which a separately trained reward model (RM) guides the generative model in learning user preferences. 
Although effective, this design inherits the pathologies of potential sampling bias that also plague reward-model-based reinforcement learning ~\cite{2025onerec,zhou2025onerecv2technicalreport}: RM training restricts sampling to a small subset of users that can only approximate global behavior distributions. It contributes to the potential information bubbles and long-tail sparsity, similar to traditional MCA. OneRec-V2~\cite{zhou2025onerecv2technicalreport} mitigates these issues by replacing proxy rewards with real user feedback signals and introducing Gradient-Bounded Policy Optimization for stable ratio clipping.

In e-commerce search, however, the feedback landscape differs fundamentally from short-video recommendation:
(a)~Unlike short-video platforms that typically present one video at a time, e-commerce search results display multiple items simultaneously. Moreover, user-item interactions (clicks, adding to cart, purchasing) follow a hierarchical progression: users typically click first, followed by subsequent actions such as adding to cart or purchasing. This contrasts sharply with video platforms where multiple interaction behaviors (like, follow, forward, dislike, comment, profile entry, etc.) can occur concurrently. Consequently, different behavioral signals in search contexts exhibit more distinct user preference patterns;
(b)~Users place greater emphasis on the strong relevance constraint between intention and the exposed item. Therefore, query-item relevance must be jointly optimized alongside conversion metrics, creating a composite reward surface that balances both relevance estimation and click probability estimation.
Meanwhile, the generated output is a discrete SID sequence ($L{=}5$ tokens) with strict hierarchical semantics (coarse→fine), where each token carries qualitatively different information. Simultaneously, for similar products with same semantics, search systems should emphasize the differentiation of unique features to provide more precise recommendations. Furthermore, Standard GRPO~\cite{shao2024grpo} assigns the same sequence-level advantage to every token, ignoring this causal structure and leading to imprecise credit assignment.

Motivated by these observations, OneSearch-V2 replaces the separately trained RM with a direct behavior feedback preference alignment system that (1)~constructs composite rewards from real user interactions, (2)~introduces a token-position marginal advantage (TPMA) mechanism for position-aware credit assignment, and (3)~supports streaming updates to handle newly emerging queries and flexible business interventions in a timely manner.

\subsubsection{Composite Reward Design}
\label{sec:reward}
We adopt GRPO as the basic optimization framework. For each rollout $o_i$ (a generated SID sequence of $L$ tokens), we compute a scalar reward $R_i$ that aggregates three complementary signals, reflecting both semantic matching quality and business conversion value.

\paragraph{Relevance Reward ($R_{\mathrm{Rel}}$).}
We leverage the existing relevance system to categorize each generated item into four tiers: 3-Excellent, 2-Related, 1-Mismatch and 0-Irrelevant. The higher level means that <query, item> pairs are more relevant.

\paragraph{Posterior Conversion Reward ($R_{\mathrm{CTR}}$).}
We utilize the calibrated posterior CTR (adaptive-weighted reward signal designed in OneSearch-v1) as a dense feedback signal. To prevent dominance of high-CTR items that may lack true relevance, the score is clipped to a bounded range (0, 1):

\paragraph{Click and Order Score ($R_{\mathrm{C\&O}}$).}
We directly reward SIDs that correspond to items the user has clicked or purchased:
\begin{equation}
    R_{\mathrm{C\&O}}(o_i) =
    \begin{cases}
        V_o, & \text{if } o_i \in \mathcal{S}_{\mathrm{order}}, \\
        V_c, & \text{if } o_i \in \mathcal{S}_{\mathrm{click}} - \mathcal{S}_{\mathrm{order}}, \\
        0, & \text{otherwise},
    \end{cases}
    \label{eq:r_co}
\end{equation}
where $\mathcal{S}_{\mathrm{order}}$ and $\mathcal{S}_{\mathrm{click}}$ denote the sets of SIDs associated with the purchased and clicked items. $v_o$ and $v_c$ are the constant reward values. This hierarchy encodes the intuition that purchases reflect stronger preference signals than clicks.

The final composite item-level reward combines them as:
\begin{equation}
    R_{\mathrm{item}}(o_i) = R_{\mathrm{C\&O}}(o_i) + R_{\mathrm{CTR}}(o_i) + R_{\mathrm{FR}}(o_i),
    \label{eq:r_rel}
\end{equation}
This additive design avoids the sparsity problem of rewards and make a well balance of relevance and conversion constrains.

\subsubsection{Standard GRPO Baseline}
\label{sec:grpo_baseline}

Group Relative Policy Optimization (GRPO) has become the dominant RL paradigm for generative retrieval systems~\cite{2025onerec,zhou2025onerecv2technicalreport,liu2025onerecthinkintextreasoninggenerative}, owing to its elimination of the critic network via within-group advantage normalization.
For each input prompt $x_u$, the current policy $\pi_\theta$ generates $G$ rollouts $\{o_i\}_{i=1}^{G}$.
The sequence-level advantage is computed as:
\begin{equation}
    \hat{A}_i = \frac{R_i - \mathrm{mean}_{j \in [G]}(R_j)}{\mathrm{std}_{j \in [G]}(R_j) + \delta},
    \label{eq:grpo_advantage}
\end{equation}
where $\delta$ is a constant for numerical stability.
The GRPO loss is:
\begin{equation}
    \mathcal{L}_{\mathrm{GRPO}} = -\frac{1}{G}\sum_{i=1}^{G}\frac{1}{|o_i|}\sum_{t=1}^{|o_i|} \min\!\Bigl(r_{i,t}\,\hat{A}_i,\; \mathrm{clip}(r_{i,t},\, 1{-}\varepsilon,\, 1{+}\varepsilon)\,\hat{A}_i\Bigr),
    \label{eq:grpo_loss}
\end{equation}
where $r_{i,t} = \pi_\theta(o_{i,t} \mid x_u, o_{i,<t}) \,/\, \pi_{\theta_{\mathrm{old}}}(o_{i,t} \mid x_u, o_{i,<t})$ is the per-token importance ratio.

\par
In the standard formulation, every token position in rollout $o_i$ receives the \emph{same} advantage $\hat{A}_i$.
However, SID generation exhibits a strict \emph{hierarchical causal structure}: the first token encodes the coarsest category, while subsequent tokens progressively refine to finer-grained attributes.
A correct first token with an incorrect second token has fundamentally different implications from the reverse.
Assigning uniform credit across positions conflates these distinct contributions and weakens the learning signal, particularly for the later, finer-grained tokens.

\subsubsection{Token-Position Marginal Advantage}
\label{sec:tpma}

To address the credit assignment limitation, we propose \textbf{TPMA-GRPO}, which decomposes the sequence-level reward into position-level marginal contributions and gates gradient flow based on prefix correctness.

\paragraph{Prefix Reward.}
For each rollout $o_i$ generating $L$ SID tokens, we define the prefix reward at position $l$ as the maximum cumulative match against any ground-truth target SID:
\begin{equation}
    R_{i,l} = \max_{t \in \mathcal{T}} \sum_{k=1}^{l} ~[o_i^k = t^k] \cdot \Delta R_{i,l}, \quad l = 1, \dots, L,
    \label{eq:prefix_reward}
\end{equation}
where $\mathcal{T} = \mathcal{S}_{\mathrm{click}} \cup \mathcal{S}_{\mathrm{order}}$ is the union of ground-truth SID sets.
$o^k_i$ and $t^k$ indicate the k-th token in rollout $o_i$ and target SID $t_i$.
This metric evaluates whether the model's generation progressively converges toward a valid target at each hierarchical level, while $\Delta R_{i,l}$ is marginal contribution at position $l$ with:
\begin{equation}
    \Delta R_{i,l} = [l < 3] \cdot 2 + [3 \le l < L] \cdot 1, \quad R_{i,0} \triangleq 0.
    \label{eq:marginal}
\end{equation}

The factor of 2 indicates the contribution of the former shared and hierarchical feature encoding (position $l < 3$) should be given more attention, compared to the latter unique feature quantization (position $3 \le l < L$). As the GR model should prioritize generating items that conform to the semantic content of the query. 0 indicates either a mismatch or that no additional match was gained.

Compared to standard GRPO, we first construct the \textit{position-level advantage} for each $l$, which normalizes marginal contributions independently across the $G$ rollouts within one group:
\begin{equation}
    \hat{A}_{i,l} = \frac{\Delta R_{i,l} - \mathrm{mean}_{j \in [G]}(\Delta R_{j,l})}{\mathrm{std}_{j \in [G]}(\Delta R_{j,l}) + \delta}.
    \label{eq:tpma_advantage}
\end{equation}
This ensures that position~$l$'s advantage is computed solely against the same position of other rollouts. Thus, each token is held responsible only for its own positional contribution, providing precise credit assignment across the coarse-to-fine hierarchy.

\paragraph{Prefix Gate.}
A critical insight is that gradient signals for latter positions are meaningful only when the prefix is correct. For example, if the former is wrong, optimizing the latter within that erroneous branch is counterproductive.
Here we introduce a prefix gate $g_{i,l}$ that modulates gradient magnitude based on prefix accuracy:

\begin{equation}
    g_{i,l} = [l=1] \cdot 1 + [l \geq 2] \cdot \frac{R_{i,l-1}}{l - 1}
    \label{eq:prefix_gate}
\end{equation}
where $g_{i,l} \in [0, 1]$.
When the prefix is perfectly matched ($R_{i,l-1} = l{-}1$), the gate is fully open ($g = 1$); when the prefix is entirely incorrect ($R_{i,l-1} = 0$), the gate closes ($g = 0$), effectively suppressing gradients for downstream tokens.
This mechanism naturally enables a hierarchical curriculum: the model first learns to generate correct coarse-level tokens before being trained on fine-grained ones.

\paragraph{Combined Advantage.}
To incorporate the richer conversion information from the item-level reward $R_{\mathrm{item}}$ (Eq.~\ref{eq:r_rel}), we first compute a group-normalized advantage:
\begin{equation}
    \hat{A}_i^{\mathrm{item}} = \frac{R_{\mathrm{item}}(o_i) - \mathrm{mean}_{j \in [G]}(R_{\mathrm{item}}(o_j))}{\mathrm{std}_{j \in [G]}(R_{\mathrm{item}}(o_j)) + \delta},
    \label{eq:seq_advantage}
\end{equation}
and combine it with the position-level one as the final advantage:
\begin{equation}
    \hat{A}_{i,l}^{\mathrm{final}} = \hat{A}_{i,l} + w_{\mathrm{item}} \cdot \hat{A}_i^{\mathrm{item}},
    \label{eq:combined_advantage}
\end{equation}
where $w_{\mathrm{seq}}$ controls the trade-off between structural prefix matching and business-oriented conversion signals.
This design allows the model to simultaneously learn \emph{what} to generate (via TPMA) and \emph{how valuable} the generation is (via the item-level reward).

\paragraph{TPMA-GRPO Loss.}
The final loss function integrates the combined advantage, prefix gate, and per-token importance ratio:
\begin{equation}
    \mathcal{L}_{\mathrm{TPMA}} = -\frac{1}{G}\sum_{i=1}^{G} \frac{1}{L}\sum_{l=1}^{L} g_{i,l} \cdot r_{i,l} \cdot \hat{A}_{i,l}^{\mathrm{final}},
    \label{eq:tpma_loss}
\end{equation}
where $r_{i,l} = \pi_\theta(o_{i,l} \mid x_u, o_{i,<l}) / \pi_{\theta_{\mathrm{old}}}(o_{i,l} \mid x_u, o_{i,<l})$ is the token-level importance ratio.
Note that we deliberately omit the clipping operation in GRPO.
The prefix gate already provides a natural regularization mechanism: when $g_{i,l} \to 0$, the effective gradient for position $l$ vanishes regardless of the ratio magnitude, preventing the gradient explosion issue.
This is analogous in spirit to GBPO proposed in OneRec-V2~\cite{zhou2025onerecv2technicalreport}, but achieves better stability through flexible structural gating rather than explicit truncation. Additional SFT is also introduced to ensure the model remain stable.

\section{Experiment}
To more thoroughly assess the effectiveness of OneSearch-V2, in this section we conduct comprehensive offline and online A/B evaluations. Moreover, extensive ablation experiments are performed to prove the feasibility of each innovations.

\textbf{Dataset and Baseline.} 
We collected the highly reliable user interactive pairs from Kuaishou's mall search platform in the past three months as the training data, and the logs from last day as the testing set. Since the V1 model has been fully deployed, all models in the offline experiments were trained using the same raw pretrained model. While for online A/B experiments, we chose the latest online OneSearch V1 as the baseline, the testing V2 is trained with the same data compared to serving version.

\textbf{Evaluation Metrics} To verify the recall and ranking performance, here we still adopt HitRate and Mean Reciprocal Ranking (MRR) as the evaluation metrics, which are widely used in search and recommendation systems. All values presented for each value were the average values for all testings.

\begin{table}[t]
\centering
\caption{Performance comparison of the proposed innovations with OneSearch on the industry dataset. The "$\backslash$+\ " means "the former model add new task", and "+" means "the sft model add new task only". The best results are in bold, and sub-optimal results are underlined in each column. }
\label{tab:offline_performance}
\begin{tabular}{p{1.8cm}cc|cc}
\toprule
\multirow{2}{*}{\textbf{Method}} & \multicolumn{2}{c|}{\textbf{Order (7229)}} & \multicolumn{2}{c}{\textbf{Click (30k)}} \\
\cmidrule(lr){2-3} \cmidrule(lr){4-5}
& \textbf{HR@10} & \textbf{MRR@10} & \textbf{HR@10} & \textbf{MRR@10}\\
\hline
OneSearch & 0.2046 & 0.0985 & 0.2231 & 0.0728 \\
$\backslash$+ CoT tasks & 0.2094 & 0.1008 & 0.2266 & 0.0731 \\
$\backslash$+ self-distill & 0.2163 & 0.1017 & 0.2398 & 0.0757 \\
$\backslash$+ rdrop & 0.2168 & 0.1045 & 0.2398 & 0.0760 \\
$\backslash$+ FGM & 0.2180 & 0.1047 & 0.2422 & 0.0766 \\
$\backslash$+ focal loss & 0.2214 & 0.1048 & 0.2471 & 0.0788 \\
\hline
+ PARS & 0.2221 & 0.1067&  \underline{0.2538} & 0.0809\\
+ GRPO & 0.2248 & 0.1106 & 0.2481 & 0.0798\\
+ TPMA & \underline{0.2265} & \underline{0.1136} & 0.2498 & \underline{0.0815} \\
\hline
OneSearch\small-V2 & \textbf{0.2314} & \textbf{0.1151} & \textbf{0.2568} & \textbf{0.0833} \\
\bottomrule
\end{tabular}
\end{table}

\begin{figure}[t]
  \centering
  \includegraphics[width=\linewidth]{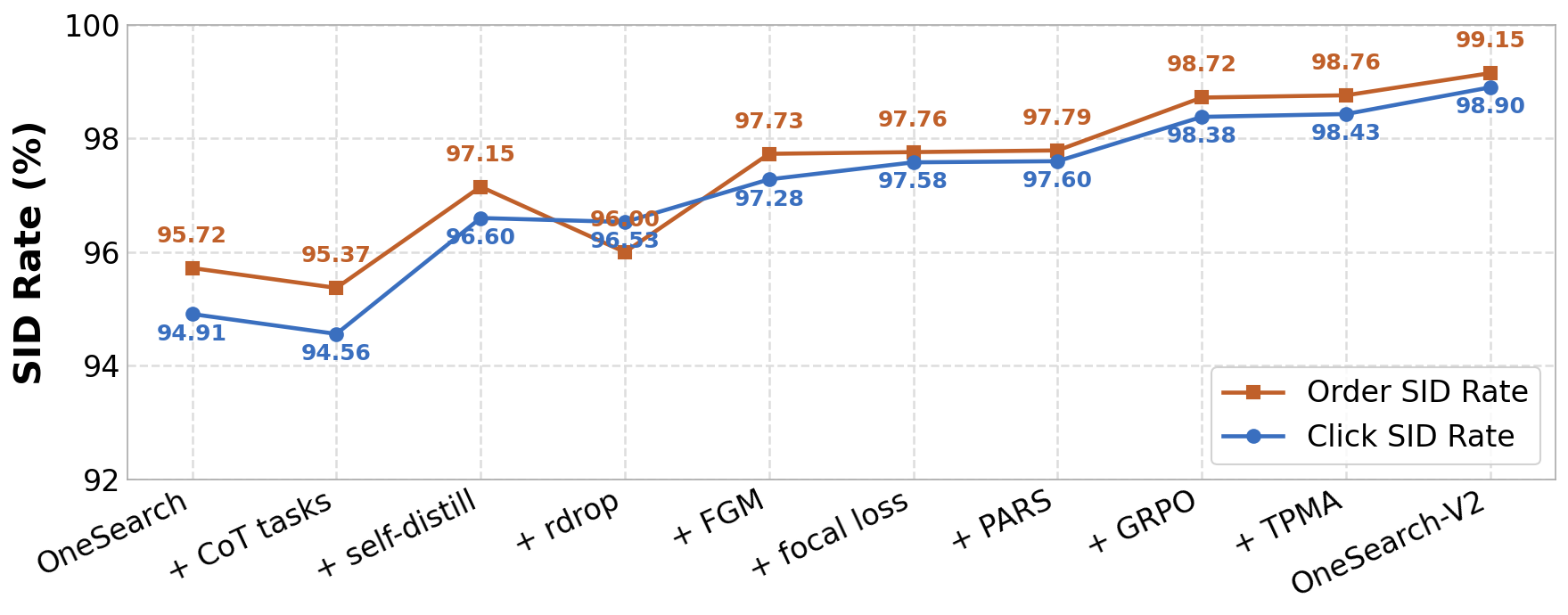}
  \caption{The sid rate of the proposed innovations with OneSearch on the industry dataset.}
  \label{fig:sid_rate}
\end{figure}

\textbf{Implementation Details}
We adopt encoder-decoder model Bart-B \cite{lewis2019bart}, decoder-only models GPT-2~\cite{brown2020} and Qwen3-0.6B~\cite{yang2025qwen3} as the base pre-trained models for the testings, in order to verify whether these innovations are applicable to different model structures. We used Qwen3-32B ~\cite{yang2025qwen3} to generate and extract the keyword-based CoT.
The beam search size is set to 512 here to strike a balance between generation quality and latency.  The batch size for SFT and DPO and GRPO is set to 512, 2048, 256, respectively, with the latter being smaller because the list-wise DPO training takes more samples as inputs. 
For the reasoning-internalized self-distillation (SFT Stage~3), we adopt the self-mode where the teacher and student share identical weights. The distillation temperature $\tau$ is set to 1.0, with the KL divergence weight $\alpha_{\text{KL}} = 0.1$ and the R-Drop coefficient $\alpha_{\text{R}} = 0.5$. For FGM adversarial training, the perturbation magnitude $\epsilon$ is set to 0.6. The focal loss parameters are set to $\alpha = 2$ and $\gamma = 3$. 
Some parameters‌ will be discussed in the following ablation study. 
$V_o$ and $V_c$ are set as 3, 4 for TPMA-GRPO. The multi-stage supervised training is conducted each, RL system is streaming training, and the keyword-base CoT generation with user interaction data is updated as close to the stream as possible. 

\subsection{Offline Performance}
We selected 30,000 page views (PVs) with valid interactions from user search logs as the testing dataset, which contains 30,000 click behaviors and 7,229 order behaviors. For each PV, we extracted the top 10 generated items to ensure a fair comparison across different methods. As shown in Table~\ref{tab:offline_performance}, the first part of our experiments aims to verify the validity of thought-augmented query understanding and reasoning-internalized self-distillation. We observe that the keyword-based CoT mechanism effectively addresses the semantic ambiguity inherent in queries. Subsequently, self-distillation further enhances the reasoning capability of OneSearch by converting deliberate, explicit CoT into inherent parameters.

The introduction of R-Drop and adversarial perturbation is also demonstrated to construct more consistent and robust predictions for each query, while the additional focal loss alleviates the extreme item class imbalance problem. Ultimately, the combinatorially optimized model achieves substantially higher recall performance (22.14\% vs. 20.46\% for order) and comparable ranking performance (10.48\% vs. 9.85\%), with an average improvement of 2.04\% in HR@10 and 0.62\% in MRR@10, compared to the baseline.

The second part of our experiments validates whether direct behavior feedback preference alignment can better meet diverse user needs without requiring a separate reward model. The adaptive reward system "+PARS" from the original OneSearch serves as our baseline. We sequentially evaluated the standard GRPO, as well as our proposed Token-Position Marginal Advantage (TPMA) mechanism. The results demonstrate that the composite reward design and the position- and item-level combined alignment formulation achieve optimal performance compared to the other methods. 

Notably, listwise DPO~\cite{sun2026listwisedpo} and GRPO~\cite{shao2024grpo} focus on complementary aspects of the optimization objective: DPO aims to refine the model's fitting of user preferences using authentic user behavioral data, while GRPO emphasizes guiding the model to generate samples that better align with reward signals through group relative optimization across multiple samples. Thus in online deployment, we first employ listwise DPO to learn the fundamental user interactive preference from real search logs, followed by TPMA to balance multiple reward objectives and enhance the model's generalization. The final version "OneSearch-V2" achieves significantly superior recall and ranking performance compared to all baseline methods. More importantly, by eliminating the dependency on separate reward models and enhancing the reasoning understanding for query content and user intent, our approach further ensures that the model can achieve healthy optimization beyond the limitations of historical logs, thereby improving both relevance and personalization in e-commerce search scenarios.

We also testing the valid SID rate for each method, which represents the proportion of valid items successfully converted among N generated SIDs. As illustrated in Fig.~\ref{fig:sid_rate}, nearly every optimization contributes to improvements in SID rate. The final OneSearch-V2, incorporating all proposed innovations, achieves optimal results (98.90\% for click, and 99.15\% for order), maintaining semantic coherence while generating diverse and relevant item candidates.

\subsection{Ablation Study}
\label{ablation}
To better examine the superiority of the proposed innovations, we evaluated that 1) the impact of query's CoT task augmentation on head and tail query, 2) the effectiveness of each part of reasoning-internalized self-distillation, and 3) Self-distillation versus alternative reasoning internalization strategies.

1) \textbf{The impact of query's CoT task augmentation on head and tail queries.}
As shown in Table~\ref{tab:ablation_head_tail_query}, the introduction of four CoT tasks into the semantic alignment procedure ($\backslash$+\ CoT tasks) yields consistent performance improvements for both head and tail query types. However, explicit CoT reasoning—wherein GR model first generates explicit CoT context before producing numerical Semantic IDs (SIDs)—significantly degrades query understanding capabilities; This finding aligns with recent studies demonstrating that explicit reasoning steps during training can adversely impact model generalization ability~\cite{yao2026compositionalgeneralizationlearnedskills}.

Incorporating keyword-based CoTs as information gain for query (+ direct CoT) at the input layer does indeed enhance overall generation performance. Nevertheless, the unbearable latency introduced by this approach renders it impractical for industrial deployment.

2) \textbf{The effectiveness of each component in reasoning internalized self-distillation.}
We isolate the contribution of each regularization technique by training it jointly with two configurations: the baseline (without self-distillation) and the self-distillation model.
As shown in Table~\ref{tab:ablation_sdft}, each technique improves the baseline independently, and self-distillation itself contributes the most substantial single improvement (+1.17\% order HR@10, +1.67\% click HR@10), confirming that internalizing keyword-guided reasoning is the primary performance driver.

\begin{table}[t]
\centering
\caption{Ablation study of reasoning-internalized self-distillation. Upper block: each technique added to the baseline; lower block: each added to the self-distillation model.}
\label{tab:ablation_sdft}
\begin{tabular}{lcccc}
\toprule
\multirow{2}{*}{\textbf{Method}}
  & \multicolumn{2}{c}{\textbf{Order (7229)}}
  & \multicolumn{2}{c}{\textbf{Click (30k)}} \\
\cmidrule(lr){2-3} \cmidrule(lr){4-5}
  & \textbf{HR@10} & \textbf{MRR@10}
  & \textbf{HR@10} & \textbf{MRR@10} \\
\midrule
Baseline
  & 0.2046 & 0.0985 & 0.2231 & 0.0728 \\
\ \ + R-Drop
  & 0.2124 & 0.1020 & 0.2292 & 0.0733 \\
\ \ + FGM
  & 0.2109 & 0.1011 & 0.2279 & 0.0732 \\
\ \ + Focal Loss
  & 0.2074 & 0.1010 & 0.2237 & 0.0723 \\
\midrule
Self-Distill
  & 0.2163 & 0.1017 & 0.2398 & 0.0757 \\
\ \ + R-Drop
  & 0.2168 & 0.1045 & 0.2398 & 0.0760 \\
\ \ + FGM
  & 0.2168 & 0.1050 & 0.2380 & 0.0757 \\
\ \ + Focal Loss
  & 0.2161 & 0.1042 & 0.2385 & 0.0753 \\
\bottomrule
\end{tabular}
\end{table}

\par
When applied on top of self-distillation, R-Drop, FGM, and focal loss each yield relatively modest individual gains. However, combining all three produces a notably larger improvement (22.14\% order HR@10 and 10.48\% MRR@10), exceeding the sum of their individual contributions. 
This observation suggests that the representation instability caused by information asymmetry between teacher and student models manifests across multiple dimensions: fragile input representations, volatile prediction outputs, and imbalanced category distributions. The synergistic effectiveness of these complementary regularization strategies indicates that they address distinct aspects of this multi-faceted instability problem. We will explore this phenomenon in greater depth in future research.

3) \textbf{Self-distillation versus alternative reasoning internalization strategies.}
To verify that self-distillation genuinely internalizes reasoning rather than merely relying on keyword input, we compare three configurations in Table~\ref{tab:ablation_teacher_student}: Base~(S), trained and evaluated without keywords; Base~(T), trained and evaluated with keywords; and the self-distilled model evaluated on both sides.
The self-distillation model Self-Distill~(S) consistently outperforms Base~(T) across all metrics, despite never observing keywords at inference time, confirming that the reasoning capability is encoded into the model weights.

Notably, before self-distillation, Base~(T) outperforms Base~(S) due to the additional keyword information; While Self-Distill~(S) slightly surpasses Self-Distill~(T). We speculate that because in self-mode distillation, the teacher and student share the same parameters, while gradients are driven entirely by the student's loss, which includes a KL constraint that encourages accurate prediction from truncated inputs. As a result, the optimization favors robustness under information-deficient conditions, enabling the student to generalize beyond the keyword-augmented teacher and achieve the best performance even without access to explicit reasoning.

\begin{table}[t]
\centering
\caption{Self-distillation model vs.\ separately trained teacher and student. ``(T)'' and ``(S)'' denote evaluation on teacher-side and student-side test data, respectively.}
\label{tab:ablation_teacher_student}
\begin{tabular}{lcccc}
\toprule
\multirow{2}{*}{\textbf{Method}}
  & \multicolumn{2}{c}{\textbf{Order (7229)}}
  & \multicolumn{2}{c}{\textbf{Click (30k)}} \\
\cmidrule(lr){2-3} \cmidrule(lr){4-5}
  & \textbf{HR@10} & \textbf{MRR@10}
  & \textbf{HR@10} & \textbf{MRR@10} \\
\midrule
Base (S)$^\dagger$
  & 0.2094 & 0.1008 & 0.2266 & 0.0731 \\
Base (T)$^\ddagger$
  & 0.2139 & 0.1011 & 0.2327 & 0.0743 \\
\midrule
Self-Distill (T)
  & 0.2155 & 0.1015 & 0.2397 & 0.0756 \\
Self-Distill (S)
  & \textbf{0.2163} & \textbf{0.1017} & \textbf{0.2398} & \textbf{0.0757} \\
\bottomrule
\end{tabular}
\begin{flushleft}
\small
$^\dagger$Student model trained and evaluated without keyword augmentation.\\
$^\ddagger$Teacher model trained and evaluated with keyword-augmented data.
\end{flushleft}
\end{table}

\par
We further compare against alternative internalization strategies in Table~\ref{tab:ablation_alternatives}. These include: (i)~special-token distillation~\cite{touvron2021}, where dedicated tokens are appended to the student input to indicate the distillation context;
(ii)~CODI-style hidden-state alignment~\cite{shen2025codi} with continuous thought vectors and L1 loss at the distillation token;
(iii)~EMA-mode~\cite{shenfeld2026selfdistillationenablescontinuallearningSDFT}, where teacher weights are an exponential moving average of the student; 
and (iv)~joint-mode~\cite{zhang2017deepmutuallearning}, where the teacher
is co-trained with the student. 
Our approach (self-mode) achieves the best performance across all metrics. Both the latent-token and alternative teacher-update strategies fall short, suggesting
that fully shared weights with input-level asymmetry is the most effective paradigm for our generative search setting.

\begin{table}[t]
\centering
\small
\caption{Comparison of alternative reasoning internalization strategies. ``Self-mode'' denotes our approach where teacher and student share identical weights.}
\label{tab:ablation_alternatives}
\begin{tabular*}{\linewidth}{@{\extracolsep{\fill}}lcccc@{}}
\toprule
\multirow{2}{*}{\textbf{Method}}
  & \multicolumn{2}{c}{\textbf{Order (7229)}}
  & \multicolumn{2}{c}{\textbf{Click (30k)}} \\
\cmidrule(lr){2-3} \cmidrule(lr){4-5}
  & \textbf{HR@10} & \textbf{MRR@10}
  & \textbf{HR@10} & \textbf{MRR@10} \\
\midrule
Baseline
  & 0.2094 & 0.1008 & 0.2266 & 0.0731 \\
\midrule
(i) Special-token
  & 0.2092 & 0.0999 & 0.2335 & 0.0739 \\
(ii) Latent + CODI
  & 0.2105 & 0.0985 & 0.2269 & 0.0714 \\
(iii) EMA-mode
  & 0.2097 & 0.1009 & 0.2317 & 0.0746 \\
(iv) Joint-mode
  & 0.2156 & 0.1016 & 0.2348 & 0.0748 \\
\midrule
Self-mode (ours)
  & \textbf{0.2163} & \textbf{0.1017} & \textbf{0.2398} & \textbf{0.0757} \\
\bottomrule
\end{tabular*}
\end{table}

\subsection{Online A/B Testing}
\label{ABTEST}
To verify OneSearch-V2’s impacts in the real online system, we compared it with the serving OneSearch-V1 in KuaiShou's mall search platform through rigorous online A/B tests. All models adopt the same deployment paradigm, means that they all take the raw query entered as input and output item candidates directly. Multiple items for the same query are generated through beam search, where an item with a higher score would be displayed firstly.

Here we trained three versions of the OneSearch-V2 model successively. OneSearch-V2\_{RAG} refers to the V1 model with additional CoT tasks in semantic alignment procedure (SFT Stage 1), and OneSearch-V2\_{Reason} further transforms User personalization learning (Stage 3) from traditional fine-tuning to self-distillation. While the final OneSearch-V2 includes all three innovations.

As shown in Table 9, all three variants of OneSearch-V2 demonstrate statistically significant improvements (P-value < 0.05) across all five key business metrics compared to the baseline OneSearch-V1 system. OneSearch-V2\_{RAG} yields consistent improvements, with Item CTR increasing by +0.52\%, PV CTR by +0.77\%, Buyer volume by +1.04\%, and Order volume by +1.07\%. These results validate that thought-augmented query understanding effectively enhances the model's capability to capture query semantics and user intent. The incorporation of reasoning-internalized self-distillation further amplifies performance gains. This demonstrates that self-distillation effectively internalizes the reasoning capabilities from the teacher model, enabling more accurate personalized predictions.

The final OneSearch-V2 model, which integrates all proposed innovations including the composite reward based position- and item-level combined alignment system, achieves the most pronounced improvements across all metrics. Specifically, it delivers +3.98\% improvement in Item CTR, +1.17\% in PV CTR, +2.90\% in PV CVR, +2.07\% in Buyer volume, and +2.11\% in Order volume. These results represent substantial conversion improvements and validate the effectiveness of our unified framework that combines thought-augmented understanding, self-distillation, and preference alignment, compare to an extra reward model.

\begin{table}[!t]
\centering
\footnotesize
\caption{Online results for A/B testing. The bold fonts means that the statistical significance (P-value) is smaller than 0.05.}
\begin{tabular*}{1\linewidth}{@{\extracolsep{\fill}}cccccc@{\extracolsep{\fill}}}
\toprule
\textbf{Method} & \textbf{Item CTR} & \textbf{PV CTR} & \textbf{PV CVR} & \textbf{Buyer} & \textbf{Order} \\
\midrule
OneSearch-V2\_{RAG} & \textbf{+0.52\%} & \textbf{+0.77\%} & \textbf{+0.63\%} & \textbf{+1.04\%} & \textbf{+1.07\%} \\
OneSearch-V2\_{Reason} & \textbf{+2.59\%} & \textbf{+1.42\%} & \textbf{+2.21\%} & \textbf{+1.50\%} & \textbf{+1.57\%} \\
\midrule
OneSearch-V2 & \textbf{+3.98\%} & \textbf{+1.17\%} & \textbf{+2.90\%} & \textbf{+2.07\%} & \textbf{+2.11\%} \\
\bottomrule
\end{tabular*}
\label{tab:online}
\end{table}

\begin{figure}[t]
  \centering
  \includegraphics[width=\linewidth]{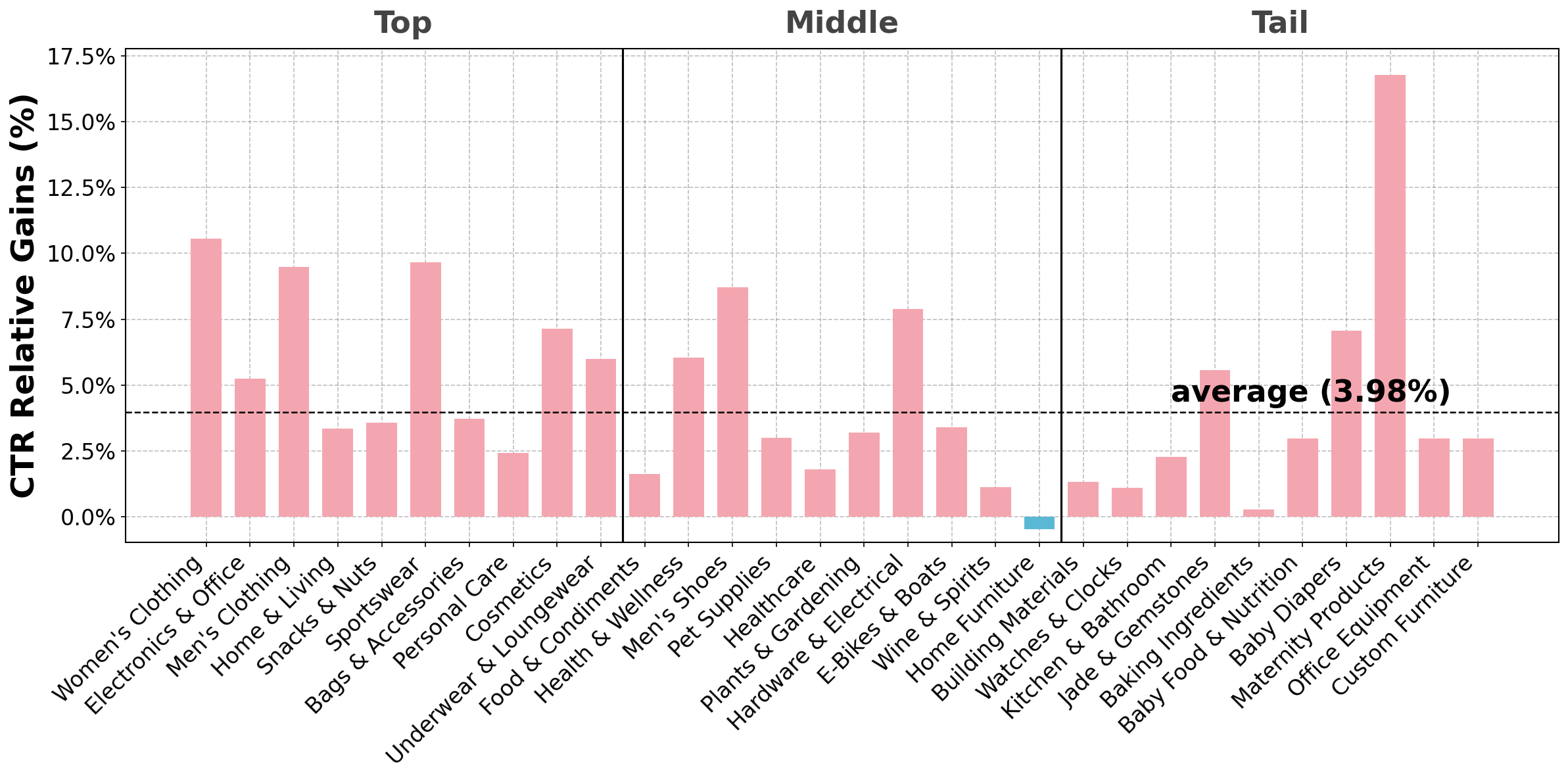}
  \caption{Relative gains in item CTR across the top/mid/tail 10 product categories ranked by page views, aggregated across multiple A/B experiment runs.}
  \label{fig:industry_ctr}
\end{figure}

We further analyzed the impact of OneSearch-V2 on the item CTR among different industries. As illustrated in Fig.~\ref{fig:industry_ctr}, we calculated the CTR relative gains across the top / middle / tail 10 industries respectively. Remarkably, almost all industries experienced increases, with an average gain of 3.98\%. These results were statistically significant, with P<0.05. Another interesting finding is that the improvements were more pronounced in categories within extensive head but ambiguous queries existing, such as Clothing, Shoes, Cosmetics, and Hardware \& Electrical, demonstrating the more accurate semantic understanding and personalized predictions of the newer model.

Last but not least, to ascertain the actual impacts on the online search experience, we conducted similar manual evaluations as OneSearch-V1. We randomly selected 200 queries and extracted 3,200 query-item pairs from identical exposure positions. We set three metrics as 1) page good rate, 2) item quality, and 3) query-item relevance. The outcomes of these assessments are presented in Table~\ref{tab:manual_evaluation}. We can see that OneSearch-V2\_Reason get the overall improvement for these metrics, and OneSearch-V2 achieves substantial increases in page good rate by 1.37\%, item quality by 0.55\%, and query item relevance by 1.65\%. The direct preference alignment can further enhances the relevance of model generation.

\begin{table}[t]
\centering
\caption{Manual evaluation results for online experience.}
\begin{tabular}{lccc}
\toprule
Metric & Page Good Rate & Item Quality & Q-I Relevance \\
\midrule
V2\_Reason & +1.12\% & +0.28\% & +1.01\% \\
V2\_Full & +1.37\% & +0.55\% & +1.65\% \\
\bottomrule
\end{tabular}
\label{tab:manual_evaluation}
\end{table}

\subsection{Further Analysis}

In this section, we mainly discuss four questions about the online deployment of the resoning enhanced OneSearch-V2 and provide our investigations to facilitate further research.

1) \textbf{What are the main aspects of online gains for OneSearch-V2?} For query frequency, we divided all queries into three categories: top queries (daily PV number larger than 1,000), middle queries (larger than 100 and less than 1,000), and long-tail queries (less than 100). For user level, we determined low-U, middle-U, and high-U based on a comprehensive analysis of user search frequency, the number of items clicked and purchased, and overall spending. For item popularity, we defined cold items as those published within the last seven days with no interaction behavior, and hot items as the top 10\% of best-selling items in each leaf category.

\begin{figure}[t]
  \centering
  \includegraphics[width=\linewidth]{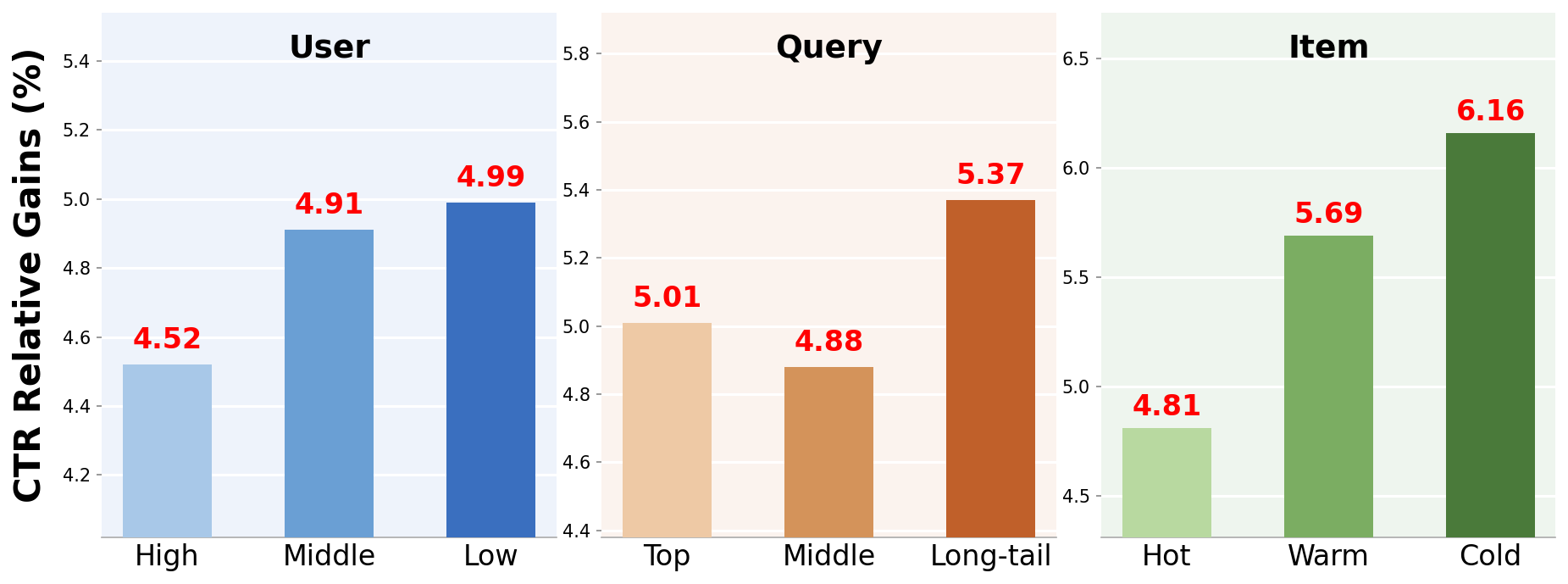}
  \caption{The CTR relative gains for various user/query/items in the most recent A/B experiment.}
  \label{fig:ctr_freq}
\end{figure}

As illustrated in the Fig.~\ref{fig:ctr_freq}, OneSearch-V2 demonstrates consistent and substantial CTR improvements across all user segments, query frequency categories, and item popularity levels, validating the robustness and generalizability of our proposed framework. Specifically, Examining the user dimension reveals particularly encouraging results for challenging user segments. And the query frequency dimension exhibits a similar trend, with long-tail queries achieving the most pronounced improvement of 5.37\%, followed by high-frequency queries at 5.01\%, and middle-frequency queries at 4.88\%. This demonstrates that CoT-enhanced semantic alignment particularly excels at handling ambiguous or rare queries where traditional systems struggle due to insufficient reasoning. 

While the item popularity analysis reveals that cold items benefit most significantly, with a remarkable 6.16\% CTR improvement, substantially outperforming warm items  at 5.69\% and hot items at 4.81\%. This finding is particularly valuable for e-commerce platforms, as effectively surfacing newly published items directly impacts merchant satisfaction and platform ecosystem health.

2) \textbf{Why Self-Distillation Outperforms Latent Token Approaches?} As shown in Table~\ref{tab:ablation_alternatives}, latent token and hidden-state alignment strategies consistently underperform our self-distillation approach.
Since CODI-style hidden-state alignment (method~ii) represents the strongest latent-token baseline, we conducted further experiments on it using the BART backbone, as reported in Table~\ref{tab:latent_comparison}. We identify two probable causes from the experimental results.

\begin{table}[t]
\centering
\small
\caption{Analysis of CODI-style configurations. ``+Proj'' adds a projection layer; ``+SD'' combines logit-level KL distillation.}
\label{tab:latent_comparison}
\begin{tabular*}{\linewidth}{@{\extracolsep{\fill}}lcccc@{}}
\toprule
\multirow{2}{*}{\textbf{Method}}
  & \multicolumn{2}{c}{\textbf{Order (7229)}}
  & \multicolumn{2}{c}{\textbf{Click (30k)}} \\
\cmidrule(lr){2-3} \cmidrule(lr){4-5}
  & \textbf{HR@10} & \textbf{MRR@10}
  & \textbf{HR@10} & \textbf{MRR@10} \\
\midrule
Baseline
  & 0.2094 & 0.1008 & 0.2266 & 0.0731 \\
Self-Distill (KL)
  & \textbf{0.2163} & \textbf{0.1017} & \textbf{0.2398} & \textbf{0.0757} \\
\midrule
CODI
  & 0.2105 & 0.0985 & 0.2269 & 0.0714 \\
CODI + Proj
  & 0.2092 & 0.0998 & 0.2270 & 0.0717 \\
CODI + Proj + SD
  & 0.2084 & 0.1002 & 0.2230 & 0.0720 \\
\bottomrule
\end{tabular*}
\end{table}

\paragraph{Supervision granularity.}
Our self-distillation provides a position-wise learning signal: at every SID token position, the student receives the teacher's full output distribution, which directly reflects how keyword information shifts the likelihoods of candidate codes.
CODI-style alignment, by contrast, supervises only a single distillation token via L1 regression of hidden activations~\cite{shen2025codi}; the remaining SID positions receive no explicit reasoning guidance.
As shown in Table~\ref{tab:latent_comparison}, adding a projection layer does not close this gap, suggesting that the limitation stems from the supervision form itself rather than model capacity.

\paragraph{Loss incompatibility.}
When we combine CODI's L1 with our logit-level KL distillation (CODI + Proj + SD in Table~\ref{tab:latent_comparison}), performance drops below either objective alone.
A plausible explanation is that the two losses impose competing constraints: L1 pulls the hidden activations toward the teacher's layer-wise geometry, while KL shapes the output distribution. The representation that best satisfies one need not best serve the other.
Our KL-only formulation sidesteps this tension, allowing the model to freely organize its internal representations around the prediction objective.

3) \textbf{Does TPMA can realize the flexible adjust for optimization objective?} 
How to conduct real-time intervention and adaptive training for dynamic optimization objectives remains a long-standing challenge for generative retrieval system. Here we conducted preliminary explorations in response to specific industrial requirements.
During the 3.18 Global Shopping Festival on the Kuaishou Platform, emerging merchants required additional traffic support to enhance their visibility and competitiveness. we implemented a targeted intervention strategy within the OneSearch-V2 framework. Specifically, for items from emerging merchants retrieved within the same query, we assigned higher relevance reward ($R_{rel}^{new}=R_{rel}^{ori}+1$). As a result, corresponding items achieved the significantly higher positions. Furthermore, higher item poster CTR values will generally result in higher rankings. This flexibility represents a significant practical advantage for industrial deployment, where business objectives frequently evolve in response to market dynamics, promotional campaigns, and strategic priorities. 

4) \textbf{What will guide further optimization for the newer OneSearch?} Future developments should be driven by three core principles: business requirements, scenario diversity, and user-centric needs. We identify several promising directions that warrant further investigation. 1) For long-tail queries with limited historical interactions, We should design more effective beyond-logs training strategies to address the insufficient sample problem. 2) E-commerce platforms increasingly feature diverse content modalities, including videos, live streams, and traditional item listings. A fundamental challenge is how to construct a unified SID tokenization scheme that can effectively represent heterogeneous content types while preserving their unique characteristics and cross-modal relationships. 3) The evolution toward agentic search systems represents another promising frontier. This paradigm shift requires innovations in efficient online learning mechanisms that can update model behavior in real-time without compromising system latency or stability.

\section{Conclusion}
This paper presents OneSearch-V2, an reasoning enhanced generative search framework addressing critical limitations in complex query understanding, personalized reasoning, and adaptive preference alignment. Through thought-augmented query understanding, reasoning-internalized self-distillation, and direct behavior feedback optimization, V2 achieves substantial improvements while maintaining deployment efficiency. Rigorous online A/B tests also demonstrate significant conversation gains, particularly for challenging queries with ambiguous semantics. The framework delivers particularly pronounced gains for challenging segments including long-tail queries, low-activity users, and cold items. 

\bibliographystyle{ACM-Reference-Format}
\bibliography{sample-base}

@String{Computing = "Computing" }

@String{Computer = "{IEEE} Computer" }

@misc{zhou2025onerecv2technicalreport,
      title={OneRec-V2 Technical Report}, 
      author={Guorui Zhou and Hengrui Hu and Hongtao Cheng and Huanjie Wang and Jiaxin Deng and Jinghao Zhang and Kuo Cai and Lejian Ren and Lu Ren and Liao Yu and Pengfei Zheng and Qiang Luo and et al.},
      year={2025},
      eprint={2508.20900},
      archivePrefix={arXiv},
      primaryClass={cs.IR},
      url={https://arxiv.org/abs/2508.20900}, 
}

@article{2024got, 
    title={Graph of Thoughts: Solving Elaborate Problems with Large Language Models}, volume={38}, 
    url={https://ojs.aaai.org/index.php/AAAI/article/view/29720}, 
    DOI={10.1609/aaai.v38i16.29720}, 
    number={16}, 
    journal={Proceedings of the AAAI Conference on Artificial Intelligence},
    author = {Besta, Maciej and Blach, Nils and Kubicek, Ales and Gerstenberger, Robert and others},
    year={2024},
    month={Mar.}, 
    pages={17682-17690} 
}

@misc{2024deepseekmath,
      title={DeepSeekMath: Pushing the Limits of Mathematical Reasoning in Open Language Models}, 
      author={Zhihong Shao and Peiyi Wang and Qihao Zhu and Runxin Xu and Junxiao Song and Xiao Bi and Haowei Zhang and Mingchuan Zhang and Y. K. Li and Y. Wu and Daya Guo},
      year={2024},
      eprint={2402.03300},
      archivePrefix={arXiv},
      primaryClass={cs.CL},
      url={https://arxiv.org/abs/2402.03300}, 
}

@inproceedings{2022cot,
 author = {Wei, Jason and Wang, Xuezhi and Schuurmans, Dale and Bosma, Maarten and ichter, brian and et al.},
 booktitle = {Advances in Neural Information Processing Systems},
 editor = {S. Koyejo and S. Mohamed and A. Agarwal and D. Belgrave and K. Cho and A. Oh},
 pages = {24824--24837},
 publisher = {Curran Associates, Inc.},
 title = {Chain-of-Thought Prompting Elicits Reasoning in Large Language Models},
 url = {https://proceedings.neurips.cc/paper_files/paper/2022/file/9d5609613524ecf4f15af0f7b31abca4-Paper-Conference.pdf},
 volume = {35},
 year = {2022}
}

@inproceedings{2023tot,
 author = {Yao, Shunyu and Yu, Dian and Zhao, Jeffrey and Shafran, Izhak and Griffiths, Tom and Cao, Yuan and Narasimhan, Karthik},
 booktitle = {Advances in Neural Information Processing Systems},
 editor = {A. Oh and T. Naumann and A. Globerson and K. Saenko and M. Hardt and S. Levine},
 pages = {11809--11822},
 publisher = {Curran Associates, Inc.},
 title = {Tree of Thoughts: Deliberate Problem Solving with Large Language Models},
 url = {https://proceedings.neurips.cc/paper_files/paper/2023/file/271db9922b8d1f4dd7aaef84ed5ac703-Paper-Conference.pdf},
 volume = {36},
 year = {2023}
}

@misc{2025onesearch,
      title={OneSearch: A Preliminary Exploration of the Unified End-to-End Generative Framework for E-commerce Search}, 
      author={Ben Chen and Xian Guo and Siyuan Wang and Zihan Liang and Yue Lv and Yufei Ma and Xinlong Xiao and et al.},
      year={2025},
      eprint={2509.03236},
      archivePrefix={arXiv},
      primaryClass={cs.IR},
      url={https://arxiv.org/abs/2509.03236}, 
}

@inproceedings{tan2024idgenrec,
    author = {Tan, Juntao and Xu, Shuyuan and Hua, Wenyue and Ge, Yingqiang and Li, Zelong and Zhang, Yongfeng},
    title = {IDGenRec: LLM-RecSys Alignment with Textual ID Learning},
    year = {2024},
    isbn = {9798400704314},
    publisher = {Association for Computing Machinery},
    address = {New York, NY, USA},
    url = {https://doi.org/10.1145/3626772.3657821},
    doi = {10.1145/3626772.3657821},
    booktitle = {Proceedings of the 47th International ACM SIGIR Conference on Research and Development in Information Retrieval},
    pages = {355–364},
    numpages = {10},
    location = {Washington DC, USA},
    series = {SIGIR '24}
}

@inproceedings{lin2025efficientinfer,
  title={Efficient Inference for Large Language Model-based Generative Recommendation},
  author={Lin, Xinyu and Yang, Chaoqun and Wang, Wenjie and Li, Yongqi and Du, Cunxiao and Feng, Fuli and Ng, See-Kiong and Chua, Tat-Seng},
  booktitle={ICLR},
  year={2025}
}

@inproceedings{yang2025earn,
    author = {Yang, Chaoqun and Lin, Xinyu and Wang, Wenjie and Li, Yongqi and Sun, Teng and Han, Xianjing and Chua, Tat-Seng},
    title = {EARN: Efficient Inference Acceleration for LLM-based Generative Recommendation by Register Tokens},
    year = {2025},
    isbn = {9798400714542},
    publisher = {Association for Computing Machinery},
    address = {New York, NY, USA},
    url = {https://doi.org/10.1145/3711896.3736919},
    doi = {10.1145/3711896.3736919},
    booktitle = {Proceedings of the 31st ACM SIGKDD Conference on Knowledge Discovery and Data Mining V.2},
    pages = {3483–3494},
    numpages = {12},
    location = {Toronto ON, Canada},
    series = {KDD '25}
}

@inproceedings{you2025r2ec,
    title={R$^2$ec: Towards Large Recommender Models with Reasoning},
    author={Runyang You and Yongqi Li and Xinyu Lin and Xin Zhang and Wenjie Wang and Wenjie Li and Liqiang Nie},
    booktitle={NeurIPS},
    year={2025}
}

@article{2025oneloc,
      title={OneLoc: Geo-Aware Generative Recommender Systems for Local Life Service}, 
      author={Zhipeng Wei and Kuo Cai and Junda She and Jie Chen and Minghao Chen and et al.},
      year={2025},
      eprint={2508.14646},
      archivePrefix={arXiv},
      primaryClass={cs.IR},
      url={https://arxiv.org/abs/2508.14646}, 
}

@article{2025onesug,
  author       = {Xian Guo and
                  Ben Chen and
                  Siyuan Wang and
                  Ying Yang and
                  Chenyi Lei and
                  et al},
  title        = {OneSug: The Unified End-to-End Generative Framework for E-commerce Query Suggestion},
  journal      = {CoRR},
  volume       = {abs/2506.06913},
  year         = {2025},
  url          = {https://doi.org/10.48550/arXiv.2506.06913},
  doi          = {10.48550/ARXIV.2506.06913},
  eprinttype    = {arXiv},
  eprint       = {2506.06913},
  bibsource    = {dblp computer science bibliography, https://dblp.org}
}

@misc{2024hstu,
      title={Actions Speak Louder than Words: Trillion-Parameter Sequential Transducers for Generative Recommendations}, 
      author={Jiaqi Zhai and Lucy Liao and Xing Liu and Yueming Wang and Rui Li and et al.},
      year={2024},
      eprint={2402.17152},
      archivePrefix={arXiv},
      primaryClass={cs.LG},
      url={https://arxiv.org/abs/2402.17152}, 
}

@misc{2025RankMixer,
      title={RankMixer: Scaling Up Ranking Models in Industrial Recommenders}, 
      author={Jie Zhu and Zhifang Fan and Xiaoxie Zhu and Yuchen Jiang and et al.},
      year={2025},
      eprint={2507.15551},
      archivePrefix={arXiv},
      primaryClass={cs.IR},
      url={https://arxiv.org/abs/2507.15551}, 
}

@inproceedings{2023tiger,
 author = {Rajput, Shashank and Mehta, Nikhil and Singh, Anima and Hulikal Keshavan, Raghunandan and Vu, Trung and et al.},
 booktitle = {Advances in Neural Information Processing Systems},
 editor = {A. Oh and T. Naumann and A. Globerson and K. Saenko and M. Hardt and S. Levine},
 pages = {10299--10315},
 publisher = {Curran Associates, Inc.},
 title = {Recommender Systems with Generative Retrieval},
 url = {https://proceedings.neurips.cc/paper_files/paper/2023/file/20dcab0f14046a5c6b02b61da9f13229-Paper-Conference.pdf},
 volume = {36},
 year = {2023}
}

@misc{2025onerec,
      title={OneRec: Unifying Retrieve and Rank with Generative Recommender and Iterative Preference Alignment}, 
      author={Jiaxin Deng and Shiyao Wang and Kuo Cai and Lejian Ren and Qigen Hu and Weifeng Ding and Qiang Luo and Guorui Zhou},
      year={2025},
      eprint={2502.18965},
      archivePrefix={arXiv},
      primaryClass={cs.IR},
      url={https://arxiv.org/abs/2502.18965}, 
}

@article{Qwen-VL,
  title={Qwen-VL: A Frontier Large Vision-Language Model with Versatile Abilities},
  author={Bai, Jinze and Bai, Shuai and Yang, Shusheng and Wang, Shijie and Tan, Sinan and Wang, Peng and Lin, Junyang and Zhou, Chang and Zhou, Jingren},
  journal={arXiv preprint arXiv:2308.12966},
  year={2023}
}

@article{yang2025qwen3,
  title={Qwen3 technical report},
  author={Yang, An and Li, Anfeng and Yang, Baosong and Zhang, Beichen and Hui, Binyuan and Zheng, Bo and Yu, Bowen and Gao, Chang and Huang, Chengen and Lv, Chenxu and others},
  journal={arXiv preprint arXiv:2505.09388},
  year={2025}
}

@article{lewis2019bart,
  title={BART: Denoising sequence-to-sequence pre-training for natural language generation, translation, and comprehension},
  author={Lewis, Mike and Liu, Yinhan and Goyal, Naman and Ghazvininejad, Marjan and Mohamed, Abdelrahman and Levy, Omer and Stoyanov, Ves and Zettlemoyer, Luke},
  journal={arXiv preprint arXiv:1910.13461},
  year={2019}
}

@article{dong2026taosr1thinkingmodelecommerce,
      title={TaoSR1: The Thinking Model for E-commerce Relevance Search}, 
      author={Chenhe Dong and Shaowei Yao and Pengkun Jiao and Jianhui Yang and Yiming Jin and Zerui Huang and Xiaojiang Zhou and Dan Ou and Haihong Tang and Bo Zheng},
      year={2026},
      journal={arXiv preprint arXiv:2508.12365},
}

@article{tang2025LREM,
      title={Large Reasoning Embedding Models: Towards Next-Generation Dense Retrieval Paradigm}, 
      author={Jianting Tang and Dongshuai Li and Tao Wen and Fuyu Lv and Dan Ou and Linli Xu},
      year={2025},
    journal={arXiv preprint arXiv:2510.14321},
}

@inproceedings{Han_2025_MTGR, 
    series={CIKM ’25},
   title={MTGR: Industrial-Scale Generative Recommendation Framework in Meituan},
   booktitle={Proceedings of the 34th ACM International Conference on Information and Knowledge Management},
   publisher={ACM},
   author={Han, Ruidong and Yin, Bin and Chen, Shangyu and Jiang, He and Jiang, Fei and Li, Xiang and Ma, Chi and Huang, Mincong and Li, Xiaoguang and Jing, Chunzhen and Han, Yueming and Zhou, MengLei and Yu, Lei and Liu, Chuan and Lin, Wei},
   year={2025},
   month=nov, pages={5731–5738},
   collection={CIKM ’25} }

@article{shen2025codi,
      title={CODI: Compressing Chain-of-Thought into Continuous Space via Self-Distillation}, 
      author={Zhenyi Shen and Hanqi Yan and Linhai Zhang and Zhanghao Hu and Yali Du and Yulan He},
      year={2025},
      journal={arXiv preprint arXiv:2502.21074},
}

@inproceedings{
    hao2025coconut,
    title={Training Large Language Models to Reason in a Continuous Latent Space},
    author={Shibo Hao and Sainbayar Sukhbaatar and DiJia Su and Xian Li and Zhiting Hu and Jason E Weston and Yuandong Tian},
    booktitle={Second Conference on Language Modeling},
    year={2025}
}

@inproceedings{
    zhang2026latentr3,
    title={Reinforced Latent Reasoning for {LLM}-based Recommendation},
    author={Yang Zhang and Wenxin Xu and Xiaoyan Zhao and Wenjie Wang and Fuli Feng and Xiangnan He and Tat-Seng Chua},
    booktitle={The Fourteenth International Conference on Learning Representations},
    year={2026}
}

@article{shenfeld2026selfdistillationenablescontinuallearningSDFT,
      title={Self-Distillation Enables Continual Learning}, 
      author={Idan Shenfeld and Mehul Damani and Jonas Hübotter and Pulkit Agrawal},
      year={2026},
      journal={arXiv preprint arXiv:2601.19897},
}

@article{zhao2026selfdistilledreasoneronpolicyselfdistillationOPSD,
      title={Self-Distilled Reasoner: On-Policy Self-Distillation for Large Language Models}, 
      author={Siyan Zhao and Zhihui Xie and Mengchen Liu and Jing Huang and Guan Pang and Feiyu Chen and Aditya Grover},
      year={2026},
      journal={arXiv preprint arXiv:2601.18734},
}

@article{SDPO,
      title={Reinforcement Learning via Self-Distillation}, 
      author={Jonas Hübotter and Frederike Lübeck and Lejs Behric and Anton Baumann and Marco Bagatella and Daniel Marta and Ido Hakimi and Idan Shenfeld and Thomas Kleine Buening and Carlos Guestrin and Andreas Krause},
      year={2026},
      journal={arXiv preprint arXiv:2601.20802},
}

@article{lixiaopeng2025genrecsurvey,
	author = {Xiaopeng Li and Bo Chen and Junda She and Shiteng Cao and You Wang and Qinlin Jia and Haiying He and Zheli Zhou and Zhao Liu and et al.},
	title = {A Survey of Generative Recommendation from a Tri-Decoupled Perspective: Tokenization, Architecture, and Optimization},
	journal = {Preprints}
}

@article{jia2025principlesapplicationscomprehensivesurvey,
      title={From Principles to Applications: A Comprehensive Survey of Discrete Tokenizers in Generation, Comprehension, Recommendation, and Information Retrieval}, 
      author={Jian Jia and Jingtong Gao and Ben Xue and Junhao Wang and Qingpeng Cai and Quan Chen and Xiangyu Zhao and Peng Jiang and Kun Gai},
      year={2025},
      journal={arXiv preprint arXiv:2502.12448},
}

@inproceedings{xiao2024BGE,
    author = {Xiao, Shitao and Liu, Zheng and Zhang, Peitian and Muennighoff, Niklas and Lian, Defu and Nie, Jian-Yun},
    title = {C-Pack: Packed Resources For General Chinese Embeddings},
    year = {2024},
    publisher = {Association for Computing Machinery},
    booktitle = {Proceedings of the 47th International ACM SIGIR Conference on Research and Development in Information Retrieval},
    pages = {641–649},
    location = {Washington DC, USA},
    series = {SIGIR '24}
}

@article{bai2025qwen3vltechnicalreport,
      title={Qwen3-VL Technical Report}, 
      author={Shuai Bai and Yuxuan Cai and Ruizhe Chen and Keqin Chen and Xionghui Chen and Zesen Cheng and Lianghao Deng and Wei Ding and Chang Gao and Chunjiang Ge and Wenbin Ge and Zhifang Guo and Qidong Huang and et al.},
      year={2025},
      journal={arXiv preprint arXiv:2511.21631},
}

@inproceedings{liang2025uniecs,
    author = {Liang, Zihan and Ma, Yufei and Qian, Zhipeng and Dai, Huangyu and Wang, Zihan and Chen, Ben and Lei, Chenyi and Ding, Yuqing and Li, Han},
    title = {UniECS: Unified Multimodal E-Commerce Search Framework with Gated Cross-modal Fusion},
    year = {2025},
    address = {New York, NY, USA},
    url = {https://doi.org/10.1145/3746252.3761170},
    doi = {10.1145/3746252.3761170},
    pages = {1788–1797},
    location = {Seoul, Republic of Korea},
    series = {CIKM '25}
}

@software{ilharco_gabriel_openclip,
  author       = {Ilharco, Gabriel and
                  Wortsman, Mitchell and
                  Wightman, Ross and
                  Gordon, Cade and
                  Carlini, Nicholas and
                  Taori, Rohan and
                  Dave, Achal and
                  Shankar, Vaishaal and
                  Namkoong, Hongseok and
                  Miller, John and
                  Hajishirzi, Hannaneh and
                  Farhadi, Ali and
                  Schmidt, Ludwig},
  title        = {OpenCLIP},
  month        = jul,
  year         = 2021,
  note         = {If you use this software, please cite it as below.},
  publisher    = {Zenodo},
  version      = {0.1},
  doi          = {10.5281/zenodo.5143773},
  url          = {https://doi.org/10.5281/zenodo.5143773}
}

@article{touvron2021,
      title={Training data-efficient image transformers \& distillation through attention}, 
      author={Hugo Touvron and Matthieu Cord and Matthijs Douze and Francisco Massa and Alexandre Sablayrolles and Hervé Jégou},
      year={2021},
      eprint={2012.12877},
      journal={https://arxiv.org/abs/2012.12877}, 
}

@article{zhang2017deepmutuallearning,
      title={Deep Mutual Learning}, 
      author={Ying Zhang and Tao Xiang and Timothy M. Hospedales and Huchuan Lu},
      year={2017},
      eprint={1706.00384},
      journal={https://arxiv.org/abs/1706.00384}, 
}

@article{brown2020,
      title={Language Models are Few-Shot Learners}, 
      author={Tom B. Brown and Benjamin Mann and Nick Ryder and Melanie Subbiah and Jared Kaplan and Prafulla Dhariwal and Arvind Neelakantan and Pranav Shyam and Girish Sastry and Amanda Askell et al.},
      year={2020},
      eprint={2005.14165},
      journal={https://arxiv.org/abs/2005.14165}, 
}

@article{yao2026compositionalgeneralizationlearnedskills,
      title={Compositional Generalization from Learned Skills via CoT Training: A Theoretical and Structural Analysis for Reasoning}, 
      author={Xinhao Yao and Ruifeng Ren and Yun Liao and Lizhong Ding and Yong Liu},
      year={2026},
      eprint={2502.04667},
      journal={arXiv preprint arXiv:2502.04667},
}

@article{sun2026listwisedpo,
      title={Listwise Direct Preference Optimization with Multi-Dimensional Preference Mixing}, 
      author={Yuhui Sun and Xiyao Wang and Zixi Li and YiTian Ding and Tianyang Ling and Jialuo Chen and Tianyi Yu and Zhenlong Yuan and Jinman Zhao},
      year={2026},
      eprint={2506.19780},
      journal={arXiv preprint arXiv:2506.19780},
}

@article{shao2024grpo,
      title={DeepSeekMath: Pushing the Limits of Mathematical Reasoning in Open Language Models}, 
      author={Zhihong Shao and Peiyi Wang and Qihao Zhu and Runxin Xu and Junxiao Song and Xiao Bi and Haowei Zhang and Mingchuan Zhang and Y. K. Li and Y. Wu and Daya Guo},
      year={2024},
      eprint={2402.03300},
      journal={arXiv preprint arXiv:2402.03300},
}

@inproceedings{NEURIPS2021Rdrop,
 author = {liang, xiaobo and Wu, Lijun and Li, Juntao and Wang, Yue and Meng, Qi and Qin, Tao and Chen, Wei and Zhang, Min and Liu, Tie-Yan},
 booktitle = {Advances in Neural Information Processing Systems},
 pages = {10890--10905},
 publisher = {Curran Associates, Inc.},
 title = {R-Drop: Regularized Dropout for Neural Networks},
 year = {2021}
}

@article{miyato2021fgm,
      title={Adversarial Training Methods for Semi-Supervised Text Classification}, 
      author={Takeru Miyato and Andrew M. Dai and Ian Goodfellow},
      year={2021},
      eprint={1605.07725},
      journal={arXiv preprint arXiv:1605.07725},
}

@article{lin2018focallossdenseobject,
      title={Focal Loss for Dense Object Detection}, 
      author={Tsung-Yi Lin and Priya Goyal and Ross Girshick and Kaiming He and Piotr Dollár},
      year={2018},
      eprint={1708.02002},
      journal={arXiv preprint arXiv:1708.02002},
}

@article{liu2025onerecthinkintextreasoninggenerative,
      title={OneRec-Think: In-Text Reasoning for Generative Recommendation}, 
      author={Zhanyu Liu and Shiyao Wang and Xingmei Wang and Rongzhou Zhang and Jiaxin Deng and Honghui Bao and Jinghao Zhang and Wuchao Li and Pengfei Zheng and Xiangyu Wu and Yifei Hu and Qigen Hu and Xinchen Luo and Lejian Ren and Zixing Zhang and Qianqian Wang and Kuo Cai and Yunfan Wu and Hongtao Cheng and Zexuan Cheng and Lu Ren and Huanjie Wang and Yi Su and Ruiming Tang and Kun Gai and Guorui Zhou},
      year={2025},
      eprint={2510.11639},
      journal={arXiv preprint arXiv:2510.11639},
}

\appendix


\section{Cross-Architecture Generalization}
\label{appendix:cross_arch}
To verify that the proposed innovations generalize across different model architectures, we conduct experiments on both GPT-2~\cite{brown2020} and Qwen3-0.6B~\cite{yang2025qwen3} (decoder-only) in addition to the BART-B (encoder-decoder) backbone used in the main paper. All models are trained on the same 5M sample dataset under comparable settings.
 
\subsection{Overall Self-Distillation Effectiveness}
\label{appendix:overall}
 
Table~\ref{tab:appendix_gpt2_ablation} and Table~\ref{tab:appendix_qwen_ablation} report the incremental results on GPT-2 and Qwen3-0.6B. Both architectures exhibit the same cumulative pattern as BART-B (Table~\ref{tab:offline_performance}), confirming the broad effectiveness of the proposed framework.
 
\begin{table}[H]
\centering
\caption{Cumulative performance of reasoning-internalized self-distillation on \textbf{GPT-2}.}
\label{tab:appendix_gpt2_ablation}
\begin{tabular}{lcccc}
\toprule
\multirow{2}{*}{\textbf{Method}}
  & \multicolumn{2}{c}{\textbf{Order (7229)}}
  & \multicolumn{2}{c}{\textbf{Click (30k)}} \\
\cmidrule(lr){2-3} \cmidrule(lr){4-5}
  & \textbf{HR@10} & \textbf{MRR@10}
  & \textbf{HR@10} & \textbf{MRR@10} \\
\midrule
Baseline
  & 0.2088 & 0.0993 & 0.2270 & 0.0733 \\
$\backslash$+ self-distill
  & 0.2128 & 0.1011 & 0.2325 & 0.0734 \\
$\backslash$+ R-Drop
  & 0.2168 & 0.1012 & 0.2380 & 0.0755 \\
$\backslash$+ FGM
  & 0.2195 & 0.1030 & 0.2430 & 0.0775 \\
$\backslash$+ focal loss
  & \textbf{0.2230} & \textbf{0.1050} & \textbf{0.2520} & \textbf{0.0802} \\
\bottomrule
\end{tabular}
\end{table}
 
\begin{table}[H]
\centering
\caption{Cumulative performance of reasoning-internalized self-distillation on \textbf{Qwen3-0.6B}.}
\label{tab:appendix_qwen_ablation}
\begin{tabular}{lcccc}
\toprule
\multirow{2}{*}{\textbf{Method}}
  & \multicolumn{2}{c}{\textbf{Order (7229)}}
  & \multicolumn{2}{c}{\textbf{Click (30k)}} \\
\cmidrule(lr){2-3} \cmidrule(lr){4-5}
  & \textbf{HR@10} & \textbf{MRR@10}
  & \textbf{HR@10} & \textbf{MRR@10} \\
\midrule
Baseline
  & 0.2195 & 0.1012 & 0.2503 & 0.0769 \\
$\backslash$+ self-distill
  & 0.2266 & 0.1060 & 0.2568 & 0.0794 \\
$\backslash$+ R-Drop
  & 0.2275 & 0.1070 & 0.2625 & 0.0800 \\
$\backslash$+ FGM
  & 0.2295 & 0.1082 & 0.2629 & 0.0809 \\
$\backslash$+ focal loss
  & \textbf{0.2310} & \textbf{0.1089} & \textbf{0.2632} & \textbf{0.0815} \\
\bottomrule
\end{tabular}
\end{table}
 
\subsection{Self-Distillation Verification}
\label{appendix:sd_verify}
 
Table~\ref{tab:appendix_gpt2_sd} and Table~\ref{tab:appendix_qwen_sd} follow the same protocol as Table~\ref{tab:ablation_teacher_student}. On both GPT-2 and Qwen3-0.6B, Self-Distill~(S) outperforms Base~(T) without observing keywords at inference, confirming architecture-agnostic reasoning internalization.
 
\begin{table}[H]
\centering
\caption{Self-distillation verification on \textbf{GPT-2}. ``(S)'' and ``(T)'' denote student-side and teacher-side evaluation.}
\label{tab:appendix_gpt2_sd}
\begin{tabular}{lcccc}
\toprule
\multirow{2}{*}{\textbf{Method}}
  & \multicolumn{2}{c}{\textbf{Order (7229)}}
  & \multicolumn{2}{c}{\textbf{Click (30k)}} \\
\cmidrule(lr){2-3} \cmidrule(lr){4-5}
  & \textbf{HR@10} & \textbf{MRR@10}
  & \textbf{HR@10} & \textbf{MRR@10} \\
\midrule
Base (S)$^\dagger$
  & 0.2088 & 0.0993 & 0.2270 & 0.0733 \\
Base (T)$^\ddagger$
  & 0.2115 & 0.1098 & 0.2298 & 0.0732 \\
\midrule
Self-Distill (T)
  & 0.2098 & 0.1002 & 0.2306 & 0.0729 \\
Self-Distill (S)
  & \textbf{0.2128} & \textbf{0.1011} & \textbf{0.2325} & \textbf{0.0734} \\
\bottomrule
\end{tabular}
\begin{flushleft}
\small
$^\dagger$Student model trained and evaluated without keyword augmentation.\\
$^\ddagger$Teacher model trained and evaluated with keyword-augmented data.
\end{flushleft}
\end{table}
 
\begin{table}[H]
\centering
\caption{Self-distillation verification on \textbf{Qwen3-0.6B}. ``(S)'' and ``(T)'' denote student-side and teacher-side evaluation.}
\label{tab:appendix_qwen_sd}
\begin{tabular}{lcccc}
\toprule
\multirow{2}{*}{\textbf{Method}}
  & \multicolumn{2}{c}{\textbf{Order (7229)}}
  & \multicolumn{2}{c}{\textbf{Click (30k)}} \\
\cmidrule(lr){2-3} \cmidrule(lr){4-5}
  & \textbf{HR@10} & \textbf{MRR@10}
  & \textbf{HR@10} & \textbf{MRR@10} \\
\midrule
Base (S)$^\dagger$
  & 0.2195 & 0.1012 & 0.2503 & 0.0769 \\
Base (T)$^\ddagger$
  & 0.2232 & 0.1035 & 0.2550 & 0.0785 \\
\midrule
Self-Distill (T)
  & 0.2241 & 0.1042 & 0.2533 & 0.0780 \\
Self-Distill (S)
  & \textbf{0.2266} & \textbf{0.1060} & \textbf{0.2568} & \textbf{0.0794} \\
\bottomrule
\end{tabular}
\begin{flushleft}
\small
$^\dagger$Student model trained and evaluated without keyword augmentation.\\
$^\ddagger$Teacher model trained and evaluated with keyword-augmented data.
\end{flushleft}
\end{table}

\section{Prompt Templates for the Reasoning Pipeline }
\label{prompt}
Table~\ref{tab:prompts} presents the complete set of prompt templates designed for the reasoning pipeline, covering three core modules: query
analysis, keyword extraction, and preference calibration. Each module
consists of a system-level role definition and a structured task prompt, collectively
enabling the understanding and condensation of complex queries.
It's worthy to note that this pipeline only applies to queries with complex intent. For other head queries, the keywords are matched by the AC automaton without introducing noise or inaccurate information during reasoning.

\definecolor{headerblue}{RGB}{52, 90, 145}
\definecolor{rowA}{RGB}{235, 240, 248}
\definecolor{rowB}{RGB}{248, 249, 251}

\newcolumntype{W}{>{\RaggedRight\arraybackslash\footnotesize}
                  p{\dimexpr\textwidth-2\tabcolsep\relax}}

\newcommand{\cell}[1]{%
  \begin{minipage}[t]{\dimexpr\textwidth-2\tabcolsep\relax}%
    \vspace{3pt}#1\vspace{5pt}%
  \end{minipage}}

\clearpage
\onecolumn

\bigskip
\setlength{\LTcapwidth}{\textwidth}
\setlength{\LTleft}{0pt}
\setlength{\LTright}{0pt}

\begin{longtable}{@{}W@{}}
  \caption{Prompt Templates for the Three-step Reasoning Pipeline}%
  \label{tab:prompts}\\
  \toprule
  \endfirsthead
  \multicolumn{1}{l}{\small\tablename~\thetable{} \textit{(continued)}}\\
  \toprule
  \endhead
  \midrule
  \multicolumn{1}{r}{\small\textit{Continued on next page}}\\
  \endfoot
  \bottomrule
  \endlastfoot

  \rowcolor{headerblue}
  \textcolor{white}{\textbf{Step 1 \;—\; Query Analysis}}\\

  \rowcolor{rowA}
  \cell{%
    \textbf{System Instruction:}\\[4pt]
    You are an AI search assistant for a Chinese e-commerce search platform.
    Analyze the user's search query across the following four dimensions.
  }\\[2pt]

  \rowcolor{rowB}
  \cell{%
    \textbf{Task Prompt:}\\[4pt]
    Analyze the query along \textbf{four dimensions}:\\[2pt]
    \textbf{1.\ Intent Underdtanding} — Identify the user's \emph{single} primary intent.%
    \begin{itemize}[leftmargin=1.5em,topsep=2pt,itemsep=0pt,parsep=0pt]
      \item \textit{Product Search}: most common (e.g.\ dress, smartphone).
      \item \textit{Functional Need}: platform features (e.g.\ track parcel).
      \item \textit{Note}: If intent $\neq$ product search, skip remaining steps.
    \end{itemize}
    \textbf{2.\ Category Identification} — Identify one or more product categories.
    \begin{itemize}[leftmargin=1.5em,topsep=2pt,itemsep=0pt,parsep=0pt]
      \item \textit{Top-level categories}: women's wear, mobile \& electronics, home goods, bags, accessories, men's wear, personal care, snacks, skincare, sports \& outdoors, cosmetics, underwear, home apparel, women's shoes, toys, gaming peripherals, fresh produce, instant food, home appliances, etc.
      \item \textit{Sub-categories}: e.g., women's wear includes T-shirts, skirts, sweatshirts, sweaters, and clothing for middle-aged and elderly women.
      \item \textit{Multiple categories}: some queries may correspond to multiple categories, e.g. ``women's windbreaker'' $\to$ women's wear \textsc{and} sports \& outdoors.
      \item \textit{Note}: Provide as comprehensive and detailed a range of product categories as possible.
    \end{itemize}
    \textbf{3.\ Attribute Recognition} — Extract attributes \emph{explicitly} stated in the query without any expansion. 
    \begin{itemize}[leftmargin=1.5em,topsep=2pt,itemsep=0pt,parsep=0pt]
      \item \textit{Common attributes}: entity, model, brand, audience,
    color, material, style, season, scene, function, price, etc.
      \item \textit{Note}: The search system must return products that match the query, so strictly retain the attributes that are relevant in the query.
    \end{itemize}
    \textbf{4.\ Topic Recommendation} — Suggest candidate topics satisfying the query, like categories or specific products.
    \begin{itemize}[leftmargin=1.5em, topsep=2pt, itemsep=0pt, parsep=0pt]
      \item \textit{Note}: need meet its categories, and attribute constraints. Do \textbf{not} over-recommend.
      \item \textit{Good cases}:
        \begin{itemize}[leftmargin=1.5em, topsep=1pt, itemsep=0pt, parsep=0pt,
                        label=$\circ$]          
          \item ``plaid skirt'' $\to$ plaid wrap skirt, plaid A-line skirt.
          \item `La Mer dupe'' $\to$ Estée Lauder serum, SK-II, Lancôme cream.
          \item ``knitwear, no turtleneck'' $\to$ V-neck knitwear, crew-neck knitwear.
          \item ``winter fruits'' $\to$ strawberry, red pomelo, orange.
        \end{itemize}
      \item \textit{Bad cases}:
        \begin{itemize}[leftmargin=1.5em, topsep=1pt, itemsep=0pt, parsep=0pt,
                        label=$\circ$]          
          \item ``bicycle accessories'' $\to$ bicycle (wrong category).
          \item ``knitwear, no turtleneck'' $\to$ turtleneck knitwear (violates constraint).
          \item ``iPhone 17'' $\to$ iPhone 16 (wrong model).
        \end{itemize}
    \end{itemize}
    Keep analysis $\leq$300 words. Please analysis query:~\texttt{\{\}}
  }\\[4pt]

  \midrule

  \rowcolor{headerblue}
  \textcolor{white}{\textbf{Step 2 \;—\; Keyword Extraction}}\\

  \rowcolor{rowA}
  \cell{%
    \textbf{System Instruction:}\\[4pt]
    You are an AI search assistant for a Chinese e-commerce search platform.
    Based on the user's search query and the LLM analysis result, extract    keywords that are \textbf{closely related} to the query.
  }\\[2pt]

  \rowcolor{rowB}
  \cell{%
    \textbf{Task Prompt:}\\[4pt]
    Rules for the extraction: \\[2pt]
    \textbf{1.\ Source Constraints:}
    \begin{itemize}[leftmargin=1.5em,topsep=2pt,itemsep=0pt,parsep=0pt]
      \item Extract \textbf{only} under ``Product Search'' intent; otherwise output \texttt{Not extractable} and stop.
      \item Extract \textbf{only} from the \textit{Topic Recommendation} section.
      \item If empty, fall back to keywords from \textit{Attribute Recognition} and \textit{Category  Identification}.
    \end{itemize}
    \textbf{2.\ Extraction Criteria:}
    \begin{itemize}[leftmargin=1.5em,topsep=2pt,itemsep=0pt,parsep=0pt]
      \item Remove off-query items (e.g.\ query ``Hisense TV'' $\Rightarrow$ exclude ``TCL TV'').
      \item Keep specific attributes (e.g.\ ``\textit{plaid} skirt'').
      \item Remove marketing terms (e.g.\ ``bestseller'', ``good quality'').
      \item Merge synonymous attributes (e.g.\ ``woolens'' merged into ``wool sweater'').
      \item Preserve model details (e.g.\ ``iPhone 15 Pro Max'').
    \end{itemize}
    \textbf{3.\ Output Format:}
    \begin{itemize}[leftmargin=1.5em,topsep=2pt,itemsep=0pt,parsep=0pt]
      \item Comma-separated; at most \textbf{8 keywords}; Each keyword can have a maximum of 10 Chinese characters.
      \item by popularity (descending).
      \item Each keyword must be independently retrievable.
    \end{itemize}
    Please extract keywords from the analysis results based on the query: \\
    \quad Query:~\texttt{\{\}} \\
    \quad Analysis Result:~\texttt{\{\}}
  }\\[4pt]

  \midrule

  \rowcolor{headerblue}
  \textcolor{white}{\textbf{Step 3 \;—\; Preference Calibration}}\\

  \rowcolor{rowA}
  \cell{%
    \textbf{System Instruction:}\\[4pt]
    You are an AI search assistant for a Chinese e-commerce search platform. Based on the user's query and behavioral history, extract or supplement  keywords from the candidate list that the user is likely interested in.
  }\\[2pt]

  \rowcolor{rowB}
  \cell{%
    \textbf{Task Prompt:}\\[4pt]
    Rules for the calibration:
    \begin{enumerate}[leftmargin=1.8em,topsep=2pt,itemsep=0pt,parsep=0pt]
      \item All keywords must be \textbf{closely related} to the user's query.
      \item Prioritize keywords aligned with the user's \textbf{profile} and
            \textbf{behavioral history} (recent searches \& recently clicked products).
      \item Prefer keywords from the candidate list; supplementation is permitted.
      \item Output at most \textbf{5 keywords}; each keyword can have a maximum of 10 Chinese characters; comma-separated;
            each keyword must be independently retrievable.
    \end{enumerate}
    Example:\\
    \textit{Input:} \\
    \quad Query: autumn-winter outfit\\
    \quad User profile: female, aged 18--23\\
    \quad Recent searches: [``autumn-winter outfit'', ``autumn-winter coat women'', ``autumn-winter trousers men'', ``hoodie'', ``fruits'']\\
    \quad Recent clicks: [``spicy hotpot instant noodles 437\,g'', ``classic rice noodles 360\,g $\times$ 8\,bags'']\\
    \quad Candidates: [``wool overcoat'', ``down jacket'',
    ``woolen coat'', ``thick hoodie'',
    ``knitwear'', ``windbreaker'', ``thermal underwear'']\\
    \textit{Output:} \\
    \quad wool overcoat women, down jacket women,
    woolen coat women, hooded hoodie women, windbreaker women\\[2pt]
    Now, based on the user's query and behavioral history, extract or supplement  keywords from the candidate list that the user is likely interested in.\\
    \textit{Input:} \\
    \quad Query:~\texttt{\{\}}\\
    \quad User profile:~\texttt{\{\}}\\
    \quad Recent searches:~\texttt{\{\}}\\
    \quad Recent clicks:~\texttt{\{\}}\\
    \quad Candidate keywords:~\texttt{\{\}}
  }\\[4pt]

\end{longtable}

\end{document}